\documentclass[aps,preprintnumbers,eqsecnum,amsmath,amssymb,nofootinbib]{revtex4}  
\usepackage{graphicx} 
\usepackage{bm}
\usepackage{slashed}
\usepackage{color}

\renewcommand\theequation{\arabic{section}.\arabic{equation}}
\begin{document}
\title{Nonfactorizable charming-loop contribution to FCNC $B_s\to \gamma l^+l^-$ decay} 
\author{Ilia Belov$^{a}$, Alexander Berezhnoy$^b$, and Dmitri Melikhov$^{b,c,d}$}
\affiliation{
$^a$INFN, Sezione di Genova, via Dodecaneso 33, I-16146 Genova, Italy\\
$^b$D.~V.~Skobeltsyn Institute of Nuclear Physics, M.~V.~Lomonosov Moscow State University, 119991, Moscow, Russia\\
$^c$Joint Institute for Nuclear Research, 141980 Dubna, Russia\\
$^d$Faculty of Physics, University of Vienna, Boltzmanngasse 5, A-1090 Vienna, Austria} 
\begin{abstract}
  We present the first theoretical calculation of nonfactorizable charm-quark loop contributions to
  the $B_s\to \gamma l^+l^-$ amplitude. We calculate the relevant form factors, $H_{A,V}^{\rm NF}(k'^2,k^2)$,
  and provide convenient parametrizations of our results
  in the form of fit functions of two variables, $k'^2$ and $k^2$, applicable in the region below
  hadron resonances, $k'^2 < M_{J/\psi}^2$ and $k^2 < M_{\phi}^2$. We report that factorizable and nonfactorizable
  charm contributions to the $B_s\to\gamma l^+l^-$ amplitude have opposite signs. To compare the charm and the
  top contributions, it is convenient to express nonfactorizable charming loop contribution
  as a non-universal (i.e., dependent on the reaction) $q^2$-dependent correction $\Delta^{\rm NF}C_7(q^2)$
  to the Wilson coefficient $C_7$. For the $B_s\to\gamma l^+l^-$ amplitude, the correction is found to be
  {\it positive}, $\Delta^{\rm NF} C_7(q^2)/C_7 > 0$. 
\end{abstract}
\date{\today}
\maketitle
\normalsize

\section{Introduction}
\label{Sec:introduction}
This paper reports the first theoretical analysis of nonfactorizable (NF) charming loops in rare
flavour-changing neutral currents (FCNC) $B_s\to\gamma l^+l^-$ decays making use of theoretical approach
formulated in \cite{bbm2023}. 

Charming loops in rare FCNC decays of the $B$-meson have visible impact on the $B$-decay observables \cite{neubert}
and their reliable theoretical description is necessary for studies of possible new physics effects (see, e.g., 
\cite{ciuchini2022,ciuchini2020,ciuchini2021,diego2021,matias2022,gubernari2022,hurth2022,stangl2022,diego2023a,diego2023b}).

A number of theoretical analyses of NF charming loops in FCNC $B$-decays has
been published in the recent years: In \cite{voloshin}, an effective gluon-photon local operator describing
the charm-quark loop has been 
calculated as an expansion in inverse charm-quark mass $m_c$ and applied to inclusive $B\to X_s\gamma$ decays
(see also \cite{ligeti,buchalla}); in \cite{khod1997}, NF corrections in $B\to K^*\gamma$ using local
operator product expansion (OPE) have been studied;
NF corrections induced by a local photon-gluon operator 
have been calculated in \cite{zwicky1,zwicky2} in terms of the light-cone (LC) 3-particle antiquark-quark-gluon
Bethe-Salpeter amplitude (3BS) of $K^*$-meson \cite{braun,ball1,ball2} with two field operators having equal coordinates,  
$\langle 0| \bar s(0)G_{\mu\nu}(0) u(x)|K^*(p)\rangle$, $x^2=0$. As noticed already long ago, local OPE for
the charm-quark loop in FCNC $B$-decays
leads to a power series in $\Lambda_{\rm QCD} m_b/m_c^2\simeq 1$.
To sum up numerically large $O(\Lambda_{\rm QCD} m_b/m_c^2)^n$ corrections, Ref.~\cite{hidr}
obtained a {\it nonlocal} photon-gluon
operator describing the charm-quark loop and evaluated its effect making use of 3BS of the $B$-meson
in a {\it collinear} LC configuration $\langle 0| \bar s(x)G_{\mu\nu}(ux) b(0)|\bar B_s(p)\rangle$, $x^2=0$
\cite{japan,braun2017}.
The same collinear approximation [known to provide the dominant 3BS contribution to meson tree-level
form factors \cite{braun1994,offen2007}] was applied also to the analysis of other FCNC $B$-decays \cite{gubernari2020}.

In later publications \cite{mk2018,m2019,m2022,m2023}, it was proven that the dominant contribution to
FCNC $B$-decay amplitudes is actually given by the convolution of a hard kernel with the 3BS in 
a different configuration --- a {\it double-collinear} light-cone configuration
$\langle 0| \bar s(y)G_{\mu\nu}(x) b(0)|\bar B_s(p)\rangle$, where $y^2=0$, $x^2=0$, but $x\,y \ne 0$. The corresponding 
factorization formula was derived in \cite{m2023}.
The first application of a double-collinear 3BS to FCNC $B_s\to\gamma\gamma$ decays was presented
in \cite{wang2022,wang2023}.

As a further step, \cite{bbm2023} developed a theoretical approach to NF charming loops in FCNC $B$-decays based
on a generic 3BS of the $B$-meson. This approach makes use of rigorous
properties of the generic 3BS:
Namely, the generic 3BS of the $B$-meson contains new Lorentz
structures (compared to the collinear and the double-collinear
configurations) and new three-particle distribution amplitudes (3DAs) that appear as the coefficients
multiplying these Lorentz structures; analyticity and continuity of the 3BS as the function of its arguments at the point
$xp=yp=x^2=y^2=0$ leads to certain constraints on the 3DAs \cite{m2023} which were implemented in the 3BS model of
\cite{bbm2023}. Moreover, \cite{bbm2023} applied this approach to $B_s\to\gamma\gamma$ decays. 

Here we extend the analysis of \cite{bbm2023} to the case of $B_s\to\gamma l^+l^-$ decays.
The paper is organized as follows:
Sect.~\ref{sec:Heff} recalls general formulas for the top contribution to the 
$B_s\to\gamma l^+l^-$ amplitude and describes the connection between the
charm contribution to the amplitudes of $B_s\to\gamma l^+l^-$ and $B_s\to\gamma^*\gamma^*$ containing two virtual
photons in the final state, including constraints on the latter imposed by electromagnetic gauge invariance. 
Section \ref{sec:charm} outlines the calculation of the factorizable and nonfactorizable charming-loop contributions to 
the $B_s\to\gamma^*\gamma^*$ amplitude.
Section \ref{sec:results} gives numerical predictions for the form factors $H_{A,V}^{\rm NF}(k'^2,k^2)$
describing NF charm in $B_s\to\gamma^*\gamma^*$ decays and compares charm contributions
with those of the top quark.
Section \ref{sec:conclusions} presents our concluding remarks.
Appendices A and B summarize some necessary details of our theoretical analysis. 
Appendices C and D contain convenient and simple fit formulas for the form factors $H_{A,V}^{\rm NF}(k'^2,k^2)$ and
$F_{TV,TA}(k'^2,k^2)$ in a broad range of their arguments $k'^2$ and $k^2$. 

\section{Top and charm contributions to $B_s\to \gamma l^+l^-$ amplitude\label{sec:Heff}}
A standard theoretical framework for the description of the FCNC $b\to s$ transitions is provided by the Wilson OPE: 
the $b\to s$ effective Hamiltonian describing dynamics at the scale $\mu$, appropriate for $B$-decays, reads 
\cite{Grinstein:1988me,Burasa,Burasb} [we use the sign convention for the effective Hamiltonian 
and the Wilson coefficients adopted in \cite{Simulaa,Simulab}]. 
\begin{eqnarray}
\label{Heff}
H_{\rm eff}^{b\to s}=\frac{G_F}{\sqrt{2}}V_{tb}V^*_{ts}\sum_i C_i(\mu) {\cal O}_i^{b\to q}(\mu), 
\end{eqnarray}
$G_F$ is the Fermi constant. The basis operators ${\cal O}_i^{b\to q}(\mu)$ contain only light degrees of freedom   
($u$, $d$, $s$, $c$, and $b$-quarks, leptons, photons and gluons); the heavy degrees of freedom of the 
SM ($W$, $Z$, and $t$-quark) are integrated out and their contributions are encoded in the Wilson coefficients
$C_i(\mu)$. 
The light degrees of freedom remain dynamical and 
the corresponding diagrams containing these particles in the loops -- in our case virtual $c$ and $u$ quarks -- 
should be calculated and added to the diagrams generated by the effective Hamiltonian.
For the SM Wilson coefficients at the scale $\mu_0=5$ GeV (the corresponding operators are listed below)
we use the recent determination [corresponding to $C_2(M_W)=-1$] from \cite{beneke2020}: 
$C_1(\mu_0)=0.147$, 
$C_2(\mu_0)=-1.053$, 
$C_7(\mu_0)=0.330$, 
$C_{9}(\mu_0)=-4.327$,
$C_{10}(\mu_0)=4.262$. 

\subsection{Top-quark contribution}
Top-quark contribution to the $\bar B_s\to \gamma l^+l^-$ amplitude is defined as follows \cite{mn2004}:  
\begin{eqnarray}
  A_{\rm top}^{(\bar B_s\to \gamma ll)}=
  \langle \gamma(q')l^+(p_1)l^-(p_2)|H_{\rm eff}^{b\to s}|\bar{B}_s(p)\rangle,\qquad q=p_1+p_2.
\end{eqnarray}
Necessary for the $\bar B_s\to\gamma l^+l^-$ decays of interest are the following terms in (\ref{Heff})\footnote{
Our notations and conventions are: 
$\gamma^5=i\gamma^0\gamma^1\gamma^2\gamma^3$, 
$\sigma_{\mu\nu}=\frac{i}{2}[\gamma_{\mu},\gamma_{\nu}]$, 
$\epsilon^{0123}=-1$, $\epsilon_{abcd}\equiv
\epsilon_{\alpha\beta\mu\nu}a^\alpha b^\beta c^\mu d^\nu$, 
$e=\sqrt{4\pi\alpha_{\rm em}}$.}:
\begin{eqnarray}
\label{b2qll}
&&H_{\rm eff}^{b\to s l^{+}l^{-}}\, =\, 
{\frac{G_{F}}{\sqrt2}}\, {\frac{\alpha_{\rm em}}{2\pi}}\, 
V_{tb}V^*_{ts}\, 
\big[-2im_b\, {\frac{C_{7}(\mu)}{q^2}}\cdot
\bar s\sigma_{\mu\nu}q^{\nu}\left (1+\gamma_5\right )b
\cdot{\bar l}\gamma^{\mu}l \nonumber\\
&&\qquad\qquad\quad +\, 
C_{9}(\mu)\cdot\bar s \gamma_{\mu}\left (1\, -\,\gamma_5 \right)   b 
\cdot{\bar l}\gamma^{\mu}l \, +\, 
C_{10}(\mu)\cdot\bar s   \gamma_{\mu}\left (1\, -\,\gamma_5 \right) b 
\cdot{\bar l}\gamma^{\mu}\gamma_{5}l \big]. 
\end{eqnarray} 
The $C_{7}$ part of $H_{\rm eff}^{b\to s l^{+}l^{-}}$ is obtained from 
\begin{eqnarray}
\label{b2qgamma}
H_{\rm eff}^{b\to s\gamma}\, &=& - \frac{G_{F}}{\sqrt2}\, V_{tb}V^*_{ts}\, 
C_{7}(\mu)\,\frac{e}{8\pi^2}\, m_b \cdot
\bar s\,\sigma_{\mu\nu}\left (1+\gamma_5\right )b \cdot F^{\mu\nu}\nonumber\\
&=& \frac{G_{F}}{\sqrt2}\, V_{tb}V^*_{ts}\, 
C_{7}(\mu)\,\frac{e}{8\pi^2}\, 2 m_b i \cdot
\bar s\,\sigma_{\mu\nu} q^\nu\left (1+\gamma_5\right )b\cdot\epsilon^\mu(q) 
\end{eqnarray}
by the replacement $\epsilon^\mu(q)\to \frac{1}{q^2} \bar l\gamma^{\mu}l\, e Q_l$, $Q_l=-1$, 
and corresponds to the diagram Fig.~\ref{Fig:Atop} (a) with the virtual photon emitted from the penguin.  
Notice that the sign of the $b\to s\gamma$ effective Hamiltonian (\ref{b2qgamma})
correlates with the sign of the electromagnetic vertex. 
For a fermion with the electric charge $Q_qe$, we use in the Feynman diagrams the vertex  
\begin{equation}
iQ_q e \bar q\gamma_\mu q \epsilon^\mu. 
\end{equation}
The $\bar B_s\to \gamma^*$ transition form factors of the basis operators in (\ref{b2qll})
are defined as \cite{mk2003} 
\begin{eqnarray}
\label{ffs}
\nonumber
\langle
\gamma(k)|\bar s \gamma_\mu\gamma_5 b|\bar B_s(p)\rangle 
&=& 
i\, e\,\varepsilon_{\alpha}(k)\, \left ( g_{\mu\alpha} \, k'k-k'_\alpha k_\mu \right )\,\frac{F_A(k'^2,k^2)}{M_{B_s}}, 
\\
\langle \gamma(k)|\bar s\gamma_\mu b|\bar B_s(p)\rangle
&=& 
e\,\varepsilon_{\alpha}(k)\,\epsilon_{\mu\alpha k' k}\frac{F_V(k'^2,k^2)}{M_{B_s}},   
\\
\langle\gamma(k)|\bar s \sigma_{\mu\nu}\gamma_5 b|\bar B_s(p) 
\rangle\, k'^{\nu}
&=& 
e\,\varepsilon_{\alpha}(k)\,\left( g_{\mu\alpha}\,k'k- k'_{\alpha}k_{\mu}\right)\, 
F_{TA}(k'^2, k^2), 
\nonumber
\\
\langle
\gamma(k)|\bar s \sigma_{\mu\nu} b|\bar B_s(p)\rangle\, k'^{\nu}
&=& 
i\, e\,\varepsilon_{\alpha}(k)\epsilon_{\mu\alpha k' k}F_{TV}(k'^2, k^2).
\nonumber 
\end{eqnarray}
We treat the form factors as functions of two variables, $F_{i}(k'^2,k^2)$: 
here $k'$ is the momentum emitted from the FCNC $b\to s$ vertex, and $k$ is the momentum 
of the (virtual) photon emitted from the valence quark of the $B$-meson, $p=k+k'$.  
The constraints on the form factors imposed by gauge invariance are discussed in
Appendix \ref{Appendix_constraints}. 
Notice that the amplitude of the operator $\bar s\sigma_{\mu\nu}b \,k'^\nu$ is reduced to a single
Lorentz structure and one form factor $F_{TA}(k'^2, k^2)$ if $k^2=0$ or $k'^2=0$.
\subsubsection{\label{sec:A1}
Direct emission of the real photon from valence quarks of the $B$ meson}

We denote as $A^{(1)}_{\rm top}$ the contribution to the $\bar B_s\to \gamma l^+l^-$ amplitude,
induced by $H_{\rm eff}^{b\to s l^+l^-}$: 
the real photon is directly emitted from the valence $s$ or $b$ quark, and the $l^+l^-$ pair
is coupled to the FCNC vertex, Fig.~\ref{Fig:Atop} (a,b). 
It corresponds to the momenta $k'=q$, $k=p-q$, $k'^2=q^2$ and $k^2=0$, and thus involves the form factors $F_i(q^2,0)$
\cite{mnk2018}:
\begin{eqnarray}
\label{A1}
A^{(1)}_{\rm top}&=&\langle\gamma (k),\,l^+(p_1),\,l^-(p_2)|H_{\rm eff}^{b\to s l^+l^-}|\bar B_s(p) \rangle\, =\,
\frac{G_F}{\sqrt{2}}\, V_{tb}V^*_{ts}\,\frac{\alpha_{\rm em}}{2\pi}\, e\, 
\varepsilon_{\alpha}(k)
\\
&& \times\left[ 
\epsilon_{\mu\alpha k' k} A_V^{(1)}(q^2)\bar l (p_2)\gamma_{\mu}l (-p_1)
-i\left (g_{\mu\alpha}\, k'k\, -\, k'_{\alpha}k_{\mu}\right)A_A^{(1)}(q^2)\bar l (p_2)\gamma_{\mu}l (-p_1)\right.\nonumber\\
&&\hspace{.2cm}\left.+\epsilon_{\mu\alpha k' k} A_{5V}^{(1)}(q^2)\bar l (p_2)\gamma_{\mu}\gamma_5 l (-p_1)
-i\left (g_{\mu\alpha}\, k'k\, -\, k'_{\alpha}k_{\mu}\right)A_{5A}^{(1)}(q^2)\bar l (p_2)\gamma_{\mu}\gamma_5 l (-p_1)
\right], \quad k'=q,\quad k=p-q.\nonumber
\end{eqnarray}
with 
\begin{eqnarray}
A_{V(A)}^{(1)}(q^2)=\frac{2\, C_{7}(\mu)}{q^2} m_bF_{TV(TA)}(q^2, 0)+C_{9}(\mu)\frac{F_{V(A)}(q^2,0)}{M_B},\quad 
A_{5V(5A)}^{(1)}(q^2)=C_{10}(\mu)\frac{F_{V(A)}(q^2,0)}{M_B}. 
\end{eqnarray}

\subsubsection{\label{sec:A2}Direct emission of the virtual photon from valence quarks of the $B$ meson}
\begin{figure}[h!]
\begin{center}
\begin{tabular}{c}   
  \includegraphics[width=16cm]{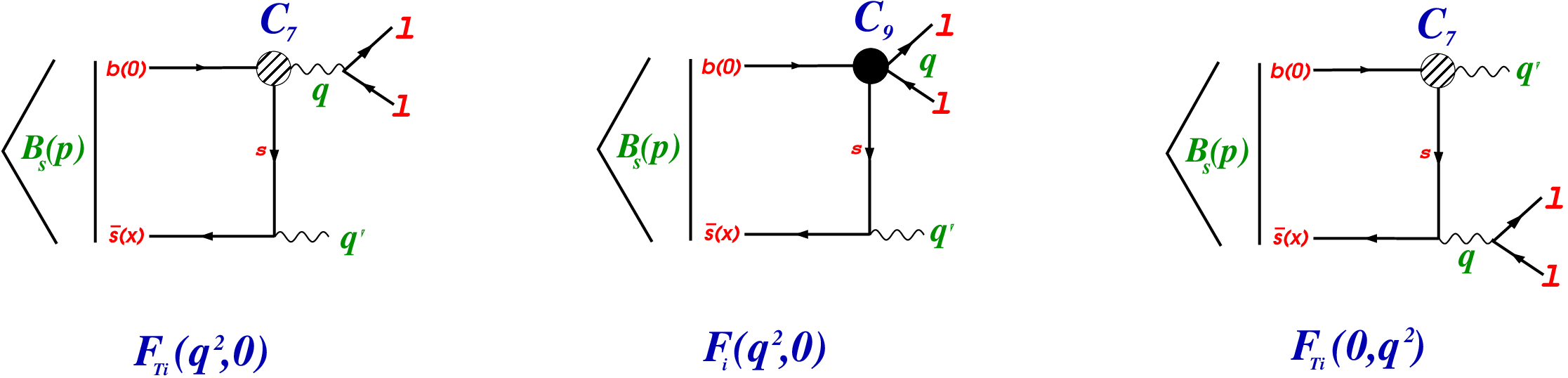} \\
  (a)  \hspace{5cm}       (b)    \hspace{5cm}       (c)\\
  \end{tabular}
\caption{Diagrams describing top-quark contributions to $\bar B_s\to \gamma l^+l^-$ amplitude.
  Dashed circle denotes the $O_7$ operator, solid circle - $O_9$ operator. Diagrams (a) and (b) describe the
  contribution $A^{(1)}$ where the real photon is emitted by spectator $s$-quark. Diagram (c) describes
  $A^{(2)}$ where the real photon is emitted from the penguin. 
  We do not show $1/m_b$-suppressed diagrams where real or virtual photon is emitted by spectator $b$-quark, see 
  \cite{mnk2018} for details.  
\label{Fig:Atop}}
\end{center}
\end{figure}

Another contribution to the amplitude, $A^{(2)}_{\rm top}$, describes the process when the real photon is emitted from the penguin FCNC vertex, whereas the virtual photon is emitted from the valence quarks of the $B$-meson,
Fig.~\ref{Fig:Atop} (c).  

The amplitude $A^{(2)}_{\rm top}$ has the same Lorentz structure as the $C_{7}$ part of $A^{(1)}$ 
where now $k=q$, $k'=p-q$, $k'^2=0$ and $k^2=q^2$. The amplitude thus involves the 
form factors $F_{TA,TV}(0,q^2)$, with $F_{TA}(0,q^2)=F_{TV}(0,q^2)$
(see details in Appendix~\ref{Appendix_constraints}): 
\begin{eqnarray}
\label{A2}
A^{(2)}_{\rm top}&=&\langle\gamma (k'),\, l^+(p_1),\,l^-(p_2)\left |H_{\rm eff}^{b\to s\gamma} \right|\bar B_s(p) \rangle\,=
\frac{G_F}{\sqrt{2}}\,V_{tb}V^*_{ts} \frac{\alpha_{\rm em}}{2\pi}\, e\,\varepsilon_{\mu}(k')\bar l (p_2)\gamma_{\alpha} l (-p_1)
\\
&\times&\left[
\epsilon_{\mu\alpha k'k}A_V^{(2)}(q^2)-i\left(g_{\mu\alpha}k'k - k'_{\alpha}k_{\mu}\right)A_A^{(2)}(q^2)
\right], 
\quad k=q, \quad k'=p-q, \nonumber
\end{eqnarray}
with 
\begin{eqnarray}
A_{V(A)}^{(2)}(q^2)=\frac{2m_b C_{7}(\mu)}{q^2}F_{TV(TA)}(0, q^2). 
\end{eqnarray}
Obviously,
\begin{eqnarray}
A^{(\bar B_s\to\gamma ll)}_{\rm top}= A^{(1)}_{\rm top}+A^{(2)}_{\rm top}.
\end{eqnarray}

\subsection{Charm-quark contribution}
\begin{figure}[b!]
\begin{center}
\begin{tabular}{c}   
  \includegraphics[width=16cm]{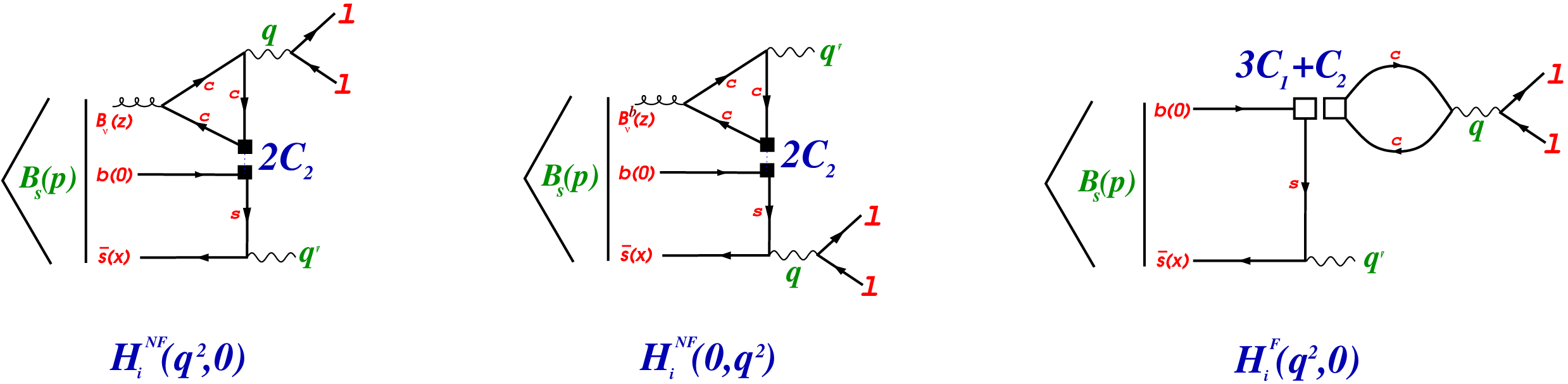}\\
  (a) \hspace{5cm} (b)  \hspace{5cm}   (c)
  \end{tabular}
\caption{The charm-loop contribution to the $B_s\to \gamma l^+l^-$ amplitude.
  (a) and (b) - nonfactorizable (NF) contributions induced by $H_{\rm eff}^{b\to s\bar cc[8\times8]}$ (solid squares),
  (c) factorizable (F) contribution induced by $H_{\rm eff}^{b\to s\bar cc[1\times1]}$ (empty squres);
  a similar factorizable contribution with the real photon emitted from the charm-quark loop vanishes and is not shown. 
  \label{Fig:Acharm}}
\end{center}
\end{figure}
The charm-loop contribution to the $B_s\to \gamma l^+l^-$ amplitude, 
\begin{equation}
  \label{AB2gll}
  A_{\rm charm}^{(\bar B_s\to\gamma ll)}=\langle \gamma(q')l^+(p_1)l^-(p_2)|H_{\rm eff}^{b\to s\bar cc}|\bar B_s(p)\rangle,
  \qquad q=p_1+p_2,
\end{equation}
is described by the diagrams of Fig.~\ref{Fig:Acharm}. $H_{\rm eff}^{b\to s\bar cc}$ includes four-quark operators
and may be written in the form
\begin{eqnarray}
\label{Hb2scc}
H_{\rm eff}^{b\to s\bar cc} &=&H_{\rm eff}^{b\to s\bar cc[1\times 1]}+H_{\rm eff}^{b\to s\bar cc[8\times 8]},\\
H_{\rm eff}^{b\to s\bar cc,[1\times1]}&=& - \frac{G_{F}}{\sqrt2}\, V_{cb}V^*_{cs}\, 
\left(C_{1}+\frac{C_2}3\right)\bar s \gamma_\mu(1-\gamma_5)b\cdot \bar c  \gamma_\mu(1-\gamma_5)c,\\
H_{\rm eff}^{b\to s\bar cc[8\times8]}&=& - \frac{G_{F}}{\sqrt2}\, V_{cb}V^*_{cs}\, 
\left(2C_2\right)\, \bar s \gamma_\mu(1-\gamma_5)t^a b \cdot \bar c  \gamma_\mu(1-\gamma_5)t^a c. 
\end{eqnarray}
Let us introduce the amplitude of the transition into two virtual photons $\gamma'$ and $\gamma$ 
\begin{eqnarray}
  \label{AB2gg}
  A_{\rm charm}^{(\bar B_s\to\gamma'\gamma)}=
 \langle \gamma'(k')\gamma(k)|H_{\rm eff}^{b\to s\bar cc}|\bar B_s(p)\rangle,\qquad p=k'+k,
\end{eqnarray}
where the photon $\gamma'(k')$ is emitted by the $c$-quark, the photon $\gamma(k)$ is emitted from the $s$-quark and 
no symmetrization over photons is performed at this point (but is done later). 
The amplitude (\ref{AB2gg}) may be written as \cite{mnk2018}
\begin{eqnarray}
  \label{Heff2b2scc}
  \langle \gamma'(k')\gamma(k)|H_{\rm eff}^{b\to s\bar cc}|\bar B_s(p)\rangle=-
  \varepsilon_\mu(k') \varepsilon_\alpha(k)H_{\mu\alpha}(k',k), 
\end{eqnarray}
with 
\begin{eqnarray}
\label{H}
H_{\mu\alpha}(k',k)=i \int dx e^{i k'x}
\langle 0| T\left\{
e Q_c \bar c(x)\gamma_\mu c(x), e Q_s \bar s(0)\gamma_\alpha s(0)\right\}|\bar B_s(p)\rangle, \qquad p=k+k'.
\end{eqnarray}
Here quark fields are understood as Heisenberg field operators with respect to all SM interactions. 
The matrix element (\ref{H}) has the Lorentz structure dictated by conservation of  
charm-quark and strange-quark vector currents that 
requires $k^\alpha H_{\mu\alpha}(k',k)=0$ and $k'^\mu H_{\mu\alpha}(k',k)=0$ (notice the absence of any contact terms): 
\begin{eqnarray}
\label{H1}
H_{\mu\alpha}(k',k)=
-\frac{G_F}{\sqrt{2}}V_{cb}V^*_{cs}e^2\left[
\epsilon_{\mu\alpha k'k}H_V
-i\left(g_{\alpha\mu}\,kk'-  k'_\alpha k_\mu\right)H_A
-i\left(k'_\alpha-\frac{kk'}{k^2}k_\alpha\right)\left(k_\mu-\frac{kk'}{k'^2}k'_\mu\right)H_3\right],  
\end{eqnarray}
with the invariant form factors $H_i$ depending on
two variables, $H_i(k'^2,k^2)$ ( $H_i$ include electric charges $Q_s$ and $Q_c$). 
The singularities in the projectors at $k^2=0$ and $k'^2=0$ should not be the singularities of the 
amplitude $H_{\mu\alpha}(k',k)$, leading to the constraints 
\begin{eqnarray}
H_3(k'^2=0,k^2)=H_3(k'^2,k^2=0)=0. 
\end{eqnarray}
As the result, $H_3$ does not contribute to the $B_s\to \gamma l^+l^-$ amplitude:  
to obtain the latter, $H_{\mu\alpha}$ should be multiplied by either $\epsilon^\alpha(k) \bar l\gamma_\mu l$ 
or $\epsilon^\mu(k') \bar l\gamma_\alpha l$. In each case, those terms in the $H_3$-part of $H_{\mu\alpha}$ 
containing $k'_\mu$ or $k'_\alpha$ vanish in the $B_s\to \gamma l^+l^-$ amplitude;
the contribution of the ``regular'' structure $k_\alpha k_\mu$ also vanishes because the form factor $H_3=0$
if $k^2=0$ or $k'^2=0$.

For the amplitude $A_{\rm charm}^{(\bar B_s\to\gamma ll)}$ we obtain  
\begin{eqnarray}
  A_{\rm charm}^{(\bar B_s\to\gamma ll)}=
  -\frac{eQ_l}{q^2}
  \left\{
  H_{\mu\alpha}(q,q')\bar l\gamma_\mu l\,\varepsilon_\alpha(q')
  +H_{\mu\alpha}(q',q)\varepsilon_\mu(q')\,\bar l\gamma_\alpha l
  \right\},\qquad Q_l=-1.
\end{eqnarray}

\subsection{Summing top and charm contributions}
Adding charm contributions to the top contributions and taking into account that $V_{tb}V_{ts}^*\simeq -V_{cb}V_{cs}^*$ leads to the following simple modifications \cite{mnk2018}: 
\begin{eqnarray}
A^{(1)}_{i}(q^2)&\to& \frac{2C_{7}}{q^2} m_b\,F_{Ti}(q^2, 0)+C_{9}\frac{F_{i}(q^2,0)}{M_B}+
{8\pi^2}\frac{H_{i}(q^2,0)}{q^2},\nonumber\\
A^{(2)}_{i}(q^2)&\to& \frac{2C_{7}}{q^2} m_b\,F_{Ti}(0, q^2)+
8\pi^2 \frac{H_{i}(0,q^2)}{q^2},\qquad i=V,A. 
\end{eqnarray} 
The full amplitude is the sum of $A^{(1)}$ and $A^{(2)}$:  
\begin{eqnarray}
\label{fullamplitude}
A_{i}(q^2)=\frac{2C_{7}}{q^2} m_b\,\left(F_{Ti}(q^2, 0)+F_{Ti}(0,q^2)\right)
 +C_{9}\frac{F_{i}(q^2,0)}{M_B}+
{8\pi^2}\frac{H_{i}(q^2,0)+H_{i}(0,q^2)}{q^2},\qquad i=V,A. 
\end{eqnarray}
The functions $H_{i}(k'^2,k^2)$ which contain factorizable and nonfactorizable charming loop contributions
will be discussed in the next Section.

\section{Charming loop contributions to $B_s\to\gamma^*\gamma^*$\label{sec:charm}}
In Eq.~(\ref{H}), quark fields are the Heisenberg operators in the SM, i.e. the corresponding $S$-matrix 
includes weak interactions of quarks. So we need to expand the S-matrix to the first order in weak interaction. 

\subsection{Factorizable contribution of the charming loop}
Factorizable contributions of the charming loop emerge in 
\begin{eqnarray}
\label{AcharmF}
{\cal A}^{(\bar B_s\to\gamma'\gamma)}_{\text{charm,F}}=
\langle \gamma'(k'),\gamma(k)|H^{b\to s\bar cc[1\times 1]}_{\rm eff}|\bar B_s(p)\rangle, 
\end{eqnarray}
when no gluons are exchanges between the charm-quark loop and the $B$-meson loop (whereas all gluon exchanges
inside the loops are allowed). The corresponding $H^{\rm F}_{\mu\alpha}(k',k)$ reads
\begin{eqnarray}
\label{HF}
H^{\rm F}_{\mu\alpha}(k',k)&=&\frac{G_F}{\sqrt2}V_{cb}V_{cs}^* \frac{3 C_1+ C_2}{3}e\,Q_c\Pi^{cc}_{\mu\nu }(k')
\left(
i\int dy e^{i k' y}
\langle 0|T\left\{
\bar s \gamma_\nu(1-\gamma_5)b(y),
e\,Q_s\,\bar s \gamma_\alpha s(0)
\right\}|\bar B_s(p)\rangle\right),  
\end{eqnarray} 
where the expression in brackets is just the amplitude of (\ref{axial-vector}) and 
\begin{eqnarray}
\label{picc1}
\Pi^{cc}_{\mu\nu}(k')=i\int dx e^{i k' x}
\langle 0|
T\{\bar c \gamma_\mu c(x),\bar c \gamma_\nu c(0)\}|0\rangle=
\left(-g_{\mu\nu} k'^2+k'_\mu k'_\nu\right)\Pi_{cc}(k'^2). 
\end{eqnarray}
For the invariant function $\Pi_{cc}(s)$ we may write the spectral representation with one subtraction 
\begin{eqnarray}
\label{picc2}
\Pi_{cc}(k'^2)=\Pi_{cc}(0)+\frac{k'^2}{\pi}\int \frac{{\rm Im}\, \Pi_{cc}(s)}{s(s-k'^2)}ds, 
\end{eqnarray}
At $k'^2\ll 4m_c^2$, $\Pi_{cc}(k'^2)$ can be calculated in perturbative QCD. At leading order in $\alpha_s$, one finds 
\begin{eqnarray}
{\rm Im}\, \Pi_{cc}(s)=\frac{N_c}{12\pi}\frac{2m_c^2+s}{s}\sqrt{1-\frac{4m_c^2}{s}}, \qquad \Pi_{cc}(0)=\frac{9}{16\pi^2}\left\{
-\frac{8}{9}\ln\left(\frac{m_c}{m_b}\right)-\frac{4}{9}\right\}.
\end{eqnarray}
The factorizable contributions to the form factors $H_i(k'^2,k^2)$ are related to $f,a_1,a_2$ (see
Appendix \ref{Appendix_constraints}) as follows 
\begin{eqnarray}
\label{HiF}
H_V^{\rm F}(k'^2,k^2)&=&\frac{3C_1+C_2}{3}Q_c \, k'^2 \Pi_{\bar cc}(k'^2) 2g(k'^2,k^2),\\
H_A^{\rm F}(k'^2,k^2)&=&\frac{3C_1+C_2}{3}Q_c \, k'^2 \Pi_{\bar cc}(k'^2)\frac{f(k'^2,k^2)}{kk'},\\
H_3^{\rm F}(k'^2,k^2)&=&\frac{3C_1+C_2}{3}Q_c \,k'^2 \Pi_{\bar cc}(k'^2)\left\{\frac{f}{kk'}+a_1+a_2\right\}.
\end{eqnarray} 
Obviously, $H_{V,A,3}^{\rm F}(k'^2=0,k^2)=0$. Therefore, the factorizable $\bar cc$ contribution
to the amplitude $A^{(2)}$ vanish;
the $\bar cc$ contributions to $A^{(2)}$ comes exclusively from NF gluon exchanges. 
The factorizable contribution to $A^{(1)}$ takes the form
[$H^{\rm F}_{3}(k'^2,k^2=0)=0$ because of the constraint (\ref{constraint1})]: 
\begin{eqnarray}
\label{HiF2}
H_i^{\rm F}(q^2,0)&=&\frac{3C_1+C_2}{3} \,Q_c \, q^2 \Pi_{\bar cc}(q^2) \frac{F_i(q^2,0)}{M_B},\quad i=A,V
\end{eqnarray}
Since $3C_1+C_2<0$ and for the $B_s\to\gamma$ transition $F_{V,A}(q^2,0)>0$ \cite{mnk2018},
we find that $H_{V,A}^{\rm F}(q^2,0)<0$ at $q^2>0$. 

Clearly, the factorizable $\bar cc$ contributions to $A^{(1)}$ can be described as a
universal $q^2$-addition to the coefficient $C_{9}$: 
\begin{eqnarray}
\label{C9eff}
C_{9}\to C^{\rm eff}_{9}(q^2)=C_{9}+\Delta^{\rm F}C_9(q^2), \quad
\Delta^{\rm F}C_9(q^2)=8\pi^2 Q_c\frac{3 C_1+C_2}{3}\Pi_{\bar cc}(q^2). 
\end{eqnarray}
Taking into account that $C_9$, $C_2$ and  $3C_1+C_2$ have the same sign, and $\Pi_{\bar cc}(q^2)\ge 0$,
we find that
\begin{eqnarray}
\label{deltaC9}
\delta^{\rm F}C_9(q^2)\equiv \Delta^{\rm F}C_9(q^2)/C_{9}>0. 
\end{eqnarray}

\subsection{Non-factorizable contribution of the charming loop}
Non-factorizable (NF) contributions of the charming loop emerge in
\begin{eqnarray}
\label{AcharmNF}
{\cal A}^{(\bar B_s\to\gamma'\gamma)}_{\text{charm,NF}}=
\langle \gamma'(k'),\gamma(k)|H^{b\to s\bar cc[8\times 8]}_{\rm eff}|\bar B_s(p)\rangle. 
\end{eqnarray}
The corresponding $H_{\mu\alpha}^{\rm NF}$ has the form 
\begin{eqnarray}
\label{HNF1}
H^{\rm NF}_{\mu\alpha}(k',k)&=&i^3\frac{G_F}{\sqrt2}V_{cb}V_{cs}^*e^2(2C_2) Q_c Q_s\int dz dx dy e^{ik' z} e^{ik y}
 \nonumber\\
 &&\times \langle 0|T\Big\{\bar c\gamma_\mu c(z),
  \bar c(0)t^a\gamma_\beta(1-\gamma_5)c(0),
  \bar c(x)t^b\gamma_\nu c(x)\Big\}|0\rangle\nonumber\\
  &&\times \langle 0|T\Big\{\bar s(y)\gamma_\alpha s(y),
  \bar s(0)t^a\gamma_\beta (1-\gamma_5)b(0)g_s B^b_\nu(x)\Big\}|\bar B_s(p)\rangle.
\end{eqnarray}
This expression takes into account photon emission by the $B$-meson valence $s$-quark; a $1/m_b$-suppressed
contribution related to photon emission by the valence $b$-quark will be omitted.
We now outline the procedure of calculating $H^{\rm NF}_{\mu\alpha}$ and for all
details refer to our recent paper \cite{bbm2023}. 


\noindent $\bullet$
The amplitude Eq.~(\ref{HNF1}) includes the charm-quark loop contribution described by the
$\langle VVA\rangle$ three-point function: 
\begin{equation}
\label{gamma_def}
\Gamma_{cc}^{\beta\nu\mu\,(ab)}\left(\kappa,k'\right) = \int dx' dz \: e^{i k' z+i \kappa x'}
\langle 0 |T\lbrace\bar c(z)\gamma^{\mu}c(z), \bar c(0)\gamma^{\beta} (1-\gamma_5)t^a c(0),
\bar c(x')\gamma^\nu t^b c(x')\rbrace|0 \rangle=\frac12\delta^{ab}\,\Gamma_{cc}^{\beta\nu\mu}(\kappa ,k'), 
\end{equation}
where $k'$ is the momentum of the external virtual photon (vertex containing index $\mu$) 
and $\kappa$ is the gluon momentum (vertex containing index $\nu$). Here $t^c$, $c=1,\dots,8$ are 
$SU_c(3)$ generators normalized as ${\rm Tr}(t^at^b)=\frac12\delta^{ab}$. 

The octet current $\bar c(0) \gamma^{\beta}(1-\gamma_5)t^a c(0)$ is a charm-quark part of the
octet-octet weak Hamiltonian. Taking into account vector-current conservation,
it is convenient to parametrize $\Gamma_{cc}^{\beta\nu\mu}(\kappa ,k')$ as follows \cite{lm} 
\begin{equation}
\label{cquarkloop}
 \Gamma_{cc}^{\beta\nu\mu}(\kappa ,k') =
-i\left(\kappa^{\beta }+k'^{\beta }\right) \epsilon^{\nu\mu\kappa k'}\,F_0
-i\left(k'^2      \epsilon^{\beta\nu\mu\kappa}-   k'^{\mu}\epsilon^{\beta\nu k'\kappa}\right)\,F_1 
-i\left(\kappa^2 \epsilon^{\beta\mu\nu k'} -\kappa^{\nu} \epsilon^{\beta\mu\kappa k'}\right)\, F_2 . 
\end{equation}
The form factors $F_{0,1,2}$ are functions of three independent invariant variables $k'^2$, $\kappa^2$, and $\kappa k'$.
We use a convenient representation of the one-loop form factors in the form \cite{bbm2023}
\begin{eqnarray}
\label{ffstriangle}
F_i\left(\kappa^2,\kappa k', k'^2\right) &=&
\frac{1}{\pi^2}\int\limits_{0}^1 d\xi \int\limits_{0}^{1-\xi}d\eta
\frac{\Delta_i(\xi,\eta)}{m_c^2 - 2\,\xi\eta\:\kappa k' - \xi(1-\xi)k'^2 - \eta(1-\eta)\kappa^2},\quad i=0,1,2,
\nonumber
\\
&&\Delta_0 = -\xi\eta, \quad \Delta_1 = \xi(1-\eta-\xi), \quad \Delta_2 = \eta(1-\eta-\xi). 
\end{eqnarray}
As shown in Sect.~III of \cite{bbm2023}, the operator describing the contribution of the charm-quark loop
may be written in the form containing only $G^a_{\nu\alpha}$:  
\begin{eqnarray}
  \int d\kappa  e^{-i\kappa x~}\Gamma_{cc}^{\beta\nu\mu\,(ab)}(\kappa ,k')B^b_{\nu}(x)dx =
 \frac{1}{4}\int d\kappa  e^{-i\kappa x~}{\overline\Gamma}_{cc}^{\beta\nu\mu\xi} (\kappa, k') G^a_{\nu\xi}(x)dx
\end{eqnarray}
with
\begin{eqnarray}
{\overline\Gamma}_{cc}^{\beta\nu\mu\xi} (\kappa, k')=
\left(\kappa ^{\beta }+k'^{\beta }\right)\epsilon^{\nu\mu\xi k'}\,F_0  
+\left(k'^{\mu}\epsilon^{\beta\nu\xi k'} + k'^2\epsilon^{\beta\nu\mu\xi}\right)\,F_1
+\left(\kappa^{\beta}\epsilon^{\xi\nu\mu k'} +
\kappa^{\mu}\epsilon^{\xi\beta\nu k'}-\kappa k'\, \epsilon^{\xi\beta\nu\mu}\right)\,F_2.     
\end{eqnarray}


\noindent $\bullet$ 
Making use of this result for the charm-quark $\langle VVA \rangle$ triangle, we have 
\begin{eqnarray}
\label{HNF2a}
H^{\rm NF}_{\mu\alpha}&=&-\frac{G_F}{\sqrt{2}}V_{cb}V^*_{cs}e^2 (2C_2) Q_sQ_c\,\tilde H^{\rm NF}_{\mu\alpha}(k',k), 
\\
\tilde H^{\rm NF}_{\mu\alpha}(k',k)&=&\frac{1}{4(2\pi)^8}\int d\tilde k dy e^{-i(\tilde k-k)y} dx d\kappa e^{-i\kappa x} 
\overline{\Gamma}_{cc}^{\beta\nu\mu\xi}(\kappa,k')
\langle 0|\bar s(y)\gamma^\eta\frac{\tilde{\slashed k}+m_s}{m_s^2-\tilde k^2}\gamma^\mu(1-\gamma^5) t^b
        G^b_{\nu\xi}(x)b(0)|\bar B_s(p)\rangle.
       \nonumber\\
\label{HNF2b}        
\end{eqnarray}
For $k'^2=0$ or $k^2=0$, $\tilde H^{\rm NF}_{\mu\alpha}$ contains 2 form factors 
\begin{eqnarray}
  \label{HNF3}
  \tilde H^{\rm NF}_{\mu\alpha}(k',k)= 
 \tilde H^{\rm NF}_V \epsilon_{\mu\alpha k' k} - i \tilde H^{\rm NF}_A \left(g_{\rho\eta}\, k'k - k_{\mu}k'_{\alpha}\right), 
\end{eqnarray}
such that 
\begin{eqnarray}
H_i^{\rm NF}(k'^2,k^2)=2C_2 Q_s Q_c \tilde H_i^{\rm NF}(k'^2,k^2). 
\end{eqnarray}
 

\noindent $\bullet$ 
The $B$-meson structure contributes to ${H}^{\rm NF}_{\mu\alpha}$ via the full set of 3BS 
\begin{eqnarray}
\langle 0|\bar s(y)\Gamma_i t^a b(0) G^a_{\nu\alpha}(x)|\bar B_s(p)\rangle,
\end{eqnarray}
with $\Gamma_i$ the appropriate combinations of $\gamma$-matrices. 
This quantity is not gauge invariant, since it contains field operators at different locations. 
To make it gauge-invariant, one needs to insert Wilson lines between the field operators.
To simplify the full consideration, it is convenient to work in a fixed-point gauge,
where the Wilson lines reduce to unity factors.

When the coordinates $x$ and $y$ are independent variables, the 3BS has the following
decomposition \cite{bbm2023}: 
\begin{eqnarray}
\label{3BSnoncoll}
&&\langle 0|\bar{s}(y)G_{\nu\xi}(x)\Gamma\, b(0)|\bar{B}_s(p)\rangle=
\frac{f_B M_B^3}{4}
\int D(\omega,\lambda)\,
e^{-i\lambda y p-i\omega x p}\, {\rm Tr}\Bigg \{\gamma_5\Gamma\,(1 +\slashed{v})
\nonumber \\
&&\times
\bigg[(p_\nu\gamma_\xi-p_\xi\gamma_\nu)\frac{1}{M_B}[\Psi_A-\Psi_V] -i\sigma_{\nu\xi}\Psi_V \nonumber\\
&&-  \frac{(x_\nu p_\xi-x_\xi p_\nu)}{x p} \left(X_A^{(x)} + \frac{\slashed{x}}{x p} M_B W^{(x)}\right)
+ \frac{(x_\nu\gamma_\xi-x_\xi\gamma_\nu)}{x p}M_B\left(Y^{(x)}_A + W^{(x)} + \frac{\slashed{x}}{x p} M_B Z^{(x)}\right)\nonumber\\
&&-  \frac{(y_\nu p_\xi-y_\xi p_\nu)}{y p} \left(X_A^{(y)} + \frac{\slashed{y}}{y p} M_B W^{(y)}\right)
+ \frac{(y_\nu\gamma_\xi-y_\xi\gamma_\nu)}{y p}M_B\left(Y^{(y)}_A + W^{(y)} + \frac{\slashed{y}}{y p} M_B Z^{(y)}\right)\nonumber\\
&&-i\epsilon_{\nu\xi\mu\beta}\,\frac{x^{\mu}p^{\beta}}{x p}\gamma^5\tilde{X}^{(x)}_A
+i\epsilon_{\nu\xi\mu\beta}\,\frac{x^{\mu}\gamma^{\beta}}{x p}\gamma^5 M_B\tilde{Y}^{(x)}_A
-i\epsilon_{\nu\xi\mu\beta}\,\frac{x^{\mu}p^{\beta}}{x p}\gamma^5\tilde{X}^{(y)}_A
+i\epsilon_{\nu\xi\mu\beta}\,\frac{x^{\mu}\gamma^{\beta}}{x p}\gamma^5 M_B\tilde{Y}^{(y)}_A  \bigg]\Bigg\},
\end{eqnarray}
where 
\begin{eqnarray}
D(\omega,\lambda)=d\omega d\lambda \theta(\omega)\theta(\lambda)\theta(1-\omega-\lambda)
\end{eqnarray}
takes into account rigorous constraints on the variables $\omega$ and $\lambda$. 
All invariant amplitudes $\Phi=\Psi_A, \Psi_V,\dots$ are functions of 5 variables, $\Phi(\omega,\lambda,x^2,y^2,xy)$, for
which we may write Taylor expansion in $x^2,y^2,xy$. Here we limit our analysis to zero-order terms in this expansion.
The corresponding zero-order terms in $\Phi$'s are functions of dimensionless arguments
$\lambda$ and $\omega$ and are referred to as the Lorentz 3DAs.

The normalization conditions for $\Psi_A$ and $\Psi_V$ have the form \cite{braun2017}:
\begin{eqnarray}
\int D(\omega,\lambda)\, \Psi_A\left(\omega,\lambda\right) = \frac{\lambda_E^2}{3M_B^2},\qquad
\int D(\omega,\lambda)\,  \Psi_V\left(\omega,\lambda\right) = \frac{\lambda_H^2}{3M_B^2}.
\end{eqnarray}
A number of Lorentz structures in (\ref{3BSnoncoll}) contain singularities at $xp=0$ and $yp=0$.
Since 3BS (\ref{3BSnoncoll}) is a continuous regular function at the point $x^2=0$, $y^2=0$, $xp=0$, $yp=0$, the absence
of singularities at $xp\to 0$ and $yp\to 0$ leads to a number of constraints on the corresponding 3DAs \cite{m2023}:   
namely, the primitives of these 3DAs should vanish at the boundaries of the 3DA support regions.
The appropriate modifications of 3DAs at the upper end-point region of $\omega$ and $\lambda$ have been
developed in \cite{bbm2023}. Here we follow the approach of \cite{bbm2023} and refer to that publication
for details. 

\noindent $\bullet$
Making use of Eq.~(\ref{3BSnoncoll})
(i) reduces the matrix element in Eq.~(\ref{HNF2b}) to trace calculation and
(ii) reduces the integrations over $x$ and $y$ to $\delta(\kappa+\omega p)\delta(\tilde k-k+\lambda p)$
(see details in \cite{bbm2023}). Using these $\delta$-functions to integrate over $\kappa$ and $k$,
the form factors $H_i$, $i={A,V}$ are obtained as integrals of the form 
\begin{eqnarray}
  \label{Hi}
  \tilde H^{\rm NF}_i(k'^2,k^2) =
  \int\limits_0^{2\omega_0} d\omega\int\limits_0^{2\omega_0-\omega} d\lambda\int\limits_0^1
  d\xi \int\limits_0^{1-\xi}d\eta\;\tilde h^{\rm NF}_i(\omega,\lambda,\xi,\eta\,|\,k'^2,k^2). 
\end{eqnarray}
Here $\tilde h_i$ are linear combinations of the 3DAs entering Eq.~(\ref{3BSnoncoll} and their primitives,
and include the form factors $F_{0,1,2}$ describing the charming $\langle VVA\rangle$ triangle 
and the $s$-quark propagator. As an illustration, we present the leading part of the $\psi_A$ and $\psi_V$ contribution to
$\tilde h_V^{\rm NF}(k'^2,k^2)$ [neglecting in the numerator all powers of $\lambda=O(\Lambda_{\rm QCD}/m_b)$
  and $\omega=O(\Lambda_{\rm QCD}/m_b)$]
\begin{eqnarray}
  \tilde h_V(\omega,\lambda,\xi,\eta\,|\,k'^2,k^2)&=&
  -\frac{\frac{1}{2}f_B M^4_B}{m_s^2+\lambda(1-\lambda)M_B^2-(1-\lambda)k^2-\lambda k'^2}\nonumber\\
&&\times\left\{
(\Psi_A+\Psi_V)\left[(M_B^2+k'^2-k^2)F_0-2k'^2 F_1\right]-2\Psi_A k'^2\left[F_0-3 F_1\right] \right\}. 
\end{eqnarray}
The form factors $F_{0,1,2}$ given by Eq.~(\ref{ffstriangle}) depend on $\xi$ and $\eta$ and should be evaluated for 
$\kappa=-\omega p$. Notice that $F_0<0$ and thus $\tilde H_V(k'^2,k^2)>0$; the form factor $\tilde H_A(k'^2,k^2)$
turns out numerically close to $\tilde H_V(k'^2,k^2)$.\footnote{
Analytic expressions for $\tilde h_i$, $i=A,V,3$ as a mathematica file may be obtained from the authors upon request.}

\section{\label{sec:results}Results for the $B_s\to \gamma l^+l^-$ NF charming loop form factors}
Our further calculation directly follows the approach of \cite{bbm2023} with the difference that
now both photons are virtual.

\subsection{Model for 3DAs}
Following \cite{bbm2023}, we make use of the set of 3DAs of local-duality (LD) model of \cite{braun2017,wang2019}
and perform the appropriate modifications of the 3DAs $X_A^{(x)}$, $X_A^{(y)}$, \ldots. All necessary details including
the explicit expressions of the 3DAs are given in Section IV of \cite{bbm2023} and will not be repeated here.
As a reference, we present just the Lorentz 3DAs $\Psi_A$ and $\Psi_V$ of the LD model \cite{braun2017}:
\begin{eqnarray}
\label{LorentzDAsviatwistDAs}
\Psi_A(\omega,\lambda) &=& (\phi_3+\phi_4)/2, \nonumber \\    
\Psi_V(\omega,\lambda) &=& (-\phi_3+\phi_4)/2 
\end{eqnarray}
with 
\begin{eqnarray}
\label{phi3}
\phi_3 &=& \frac{105(\lambda_E^2-\lambda_H^2)}{32\omega_0^7M_B^2}\,
\lambda\omega^2\left(2\omega_0-\omega-\lambda\right)^2\theta\left(2\omega_0-\omega-\lambda\right),\\
  \label{phi4}
  \phi_4 &=& \frac{35(\lambda_E^2+\lambda_H^2)}{32\omega_0^7M_B^2}\,
  \omega^2\left(2\omega_0-\omega-\lambda\right)^3\theta\left(2\omega_0-\omega-\lambda\right). 
\end{eqnarray} 
Dimensionless parameter $\omega_0$ is related to $\lambda_{B}$, the inverse moment of the $B$-meson LC
distribution amplitude, as 
\begin{eqnarray}
  \label{omega0}
\omega_0=\frac{5}{2}\frac{\lambda_{B}}{M_{B}}. 
\end{eqnarray}
For this model, the integration limits take the following form ($2\omega_0<1$): 
\begin{eqnarray}
  \int D(\omega,\lambda)\,\theta\left(2\omega_0-\omega-\lambda\right)\,{(\dots)}=
  \int\limits_0^{2\omega_0} d \omega \int\limits_0^{2\omega_0-\omega}d\lambda\,{(\dots)}. 
\end{eqnarray}
The form factors $H_{A,V}(k'^2,k^2)$ have explicit linear dependence on $\lambda_{E,H}^2$, so we write  
\begin{eqnarray}
  H^{\rm NF}_{i}(k'^2,k^2)=2C_2 Q_s Q_c
  \left[R_{iE}(k'^2,k^2)\lambda_E^2+R_{iH}(k'^2,k^2)\lambda_H^2\right], \qquad i=A,V.
\end{eqnarray}
QCD sum rules suggest an approximate relation \cite{braun2017} 
\begin{eqnarray}
\label{sr4lambdas}
\lambda_H^2 \simeq 2\lambda_E^2.
\end{eqnarray}
Then, the appropriate combinations of the form factors which describe NF charm contributions have the form 
\begin{eqnarray}
  R_i(k'^2,k^2)=R_{iE}(k'^2,k^2)+2R_{iH}(k^2,k^2),\qquad
  H_{i}(k'^2,k^2)=2C_2 Q_s Q_c \lambda_E^2 R_{i}(k'^2,k^2), \qquad i=A,V.
\end{eqnarray}
Combining (\ref{sr4lambdas}) with QCD equations of motion,
3BS model used in our analysis leads to approximate relations
\begin{eqnarray}
\label{lambdaEH}
\lambda_H^2 \simeq 1.2\lambda_{B_s}^2, \qquad \lambda_E^2 \simeq 0.6\lambda_{B_s}^2. 
\end{eqnarray}
This leads to an explicit linear dependence of $H_i$ on $\lambda_{B_s}^2$. 
It should be noticed however that the form factors 
have also a complicated implicit dependence on $\lambda_{B_s}$ through a $\lambda_{B_s}$-dependent shape of the
three-particle distribution amplitudes of $B_s$.
We present below our results for the benchmark point \cite{bbm2023}
\begin{eqnarray}
\label{lambdaBs}
\lambda_{B_s}(1\; {\rm GeV})=0.45\; {\rm GeV}.
\end{eqnarray}
For a discussion of the existing estimates of $\lambda_{B_s}$ and $\lambda^2_{E,H}$
including their dependence on the scale, we refer to
\cite{khodjamirian2020,kou,braun2004,beneke2011,wang2016,wang2018,wang2021,zwicky2021,im2022,wang2022a}.
A remark may be useful before we go to the results for the form factors:
The parameters $\lambda^2_{E,H}$ as well as the parameter $\lambda_{B_s}$ depend on the scale. We do not discuss
this dependence and perform numerical calculations just assuming that the form factors are represented in terms
of these parameters at a fixed scale 1 GeV as given in (\ref{lambdaEH}) and (\ref{lambdaBs}).
The parameter $\lambda_{B_s}$ is presently not known with good accuracy (see \cite{im2022}) 
and at the same time one observes a sizeable
sensitivity of the form factors $H^{\rm NF}_i$ to its value (see Fig.~7 in \cite{bbm2023} for the
form factor $H^{\rm NF}_i(0,0)$).
In this situation, studying the scale-dependence of the form factors $H^{\rm NF}_i(k'^2,k^2)$ induced by the
scale-dependence of $\lambda_{B_s}$ would not be useful. 
  
\subsection{Results for the form factors $H^{\rm NF}_i(k'^2,k^2)$}
The analytic expressions (\ref{Hi}) are based on finite-order QCD diagrams and thus
cannot be trusted near quark thresholds. For instance, the calculated form factors exhibit
steep rise at negative $k^2\to 0$ which is unphysical, as the nearest hadron pole lies at $k^2=M_\phi^2$ and  
the two-meson threshold lies at $k^2=4M_K^2$.

We therefore pursue the strategy previously applied in \cite{ms2000,bbm2023} to the form factors of one variable, 
and extend it to the case of $R_i(k'^2,k^2)$ depending on two variables:
We first calculate the form factors $R_i(k'^2,k^2)$ using the analytic
expressions (\ref{Hi}) in the rectangular region relatively far from quark thresholds in QCD diagrams: 
\begin{eqnarray}
  \label{rectangular}
\qquad 0 <k'^2({\rm GeV}^2) < 4, \quad  -5 <k^2({\rm GeV}^2) < -0.6. 
\end{eqnarray}
We then {\it interpolate} the results obtained in this region as function of two variables $k'^2$ and $k^2$
by a formula (\ref{fitHVNF}) which takes into account the correct location of the hadron poles at
$k'^2=M_{J/\psi}^2$ and $k^2=M_{\phi}^2$ and contains a number of fit parameters allowing to fit $R_i(k'^2,k^2)$
in the rectangular region (\ref{rectangular}) with an accuracy of not worse than 2 percent.
The fit formula and its parameters obtained by this procedure are given in Appendix \ref{Appendix:C}. 
Finally, since our interpolating formula takes into account correct location of the lowest meson poles
in the $k'^2$ and $k^2$ channels, we find it eligible to use this formula
to {\it extrapolate} the form factors to the region of timelike momenta $k'^2\le M^2_{J/\psi}$ and $k^2\le 4M_K^2$.

\begin{figure}[th!]
\begin{center}
\begin{tabular}{cc}   
  \includegraphics[height=5cm]{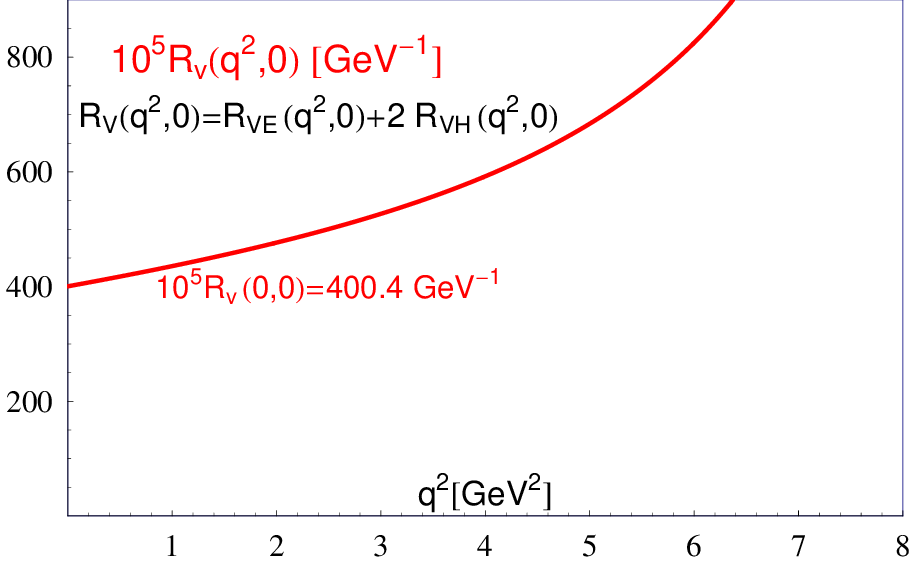}     &  \includegraphics[height=5cm]{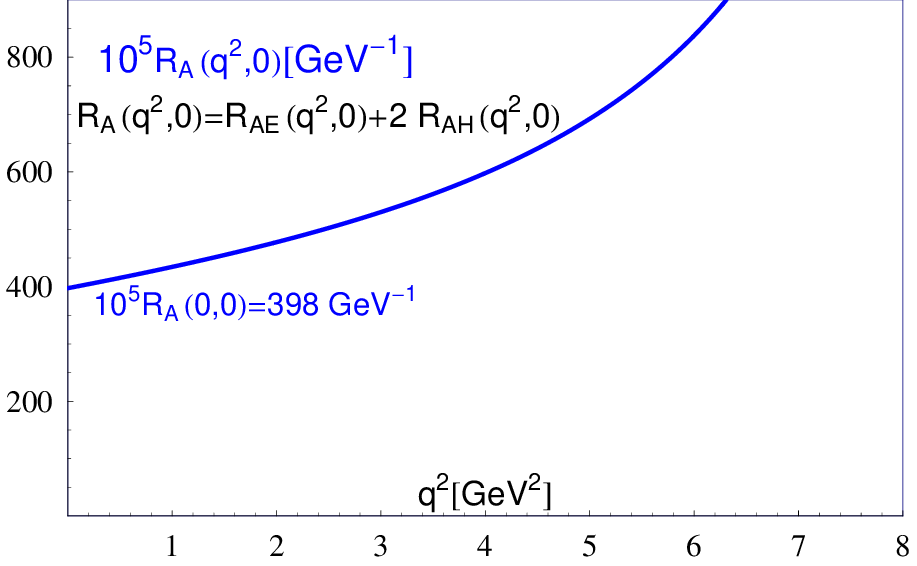}\\
   \includegraphics[height=5cm]{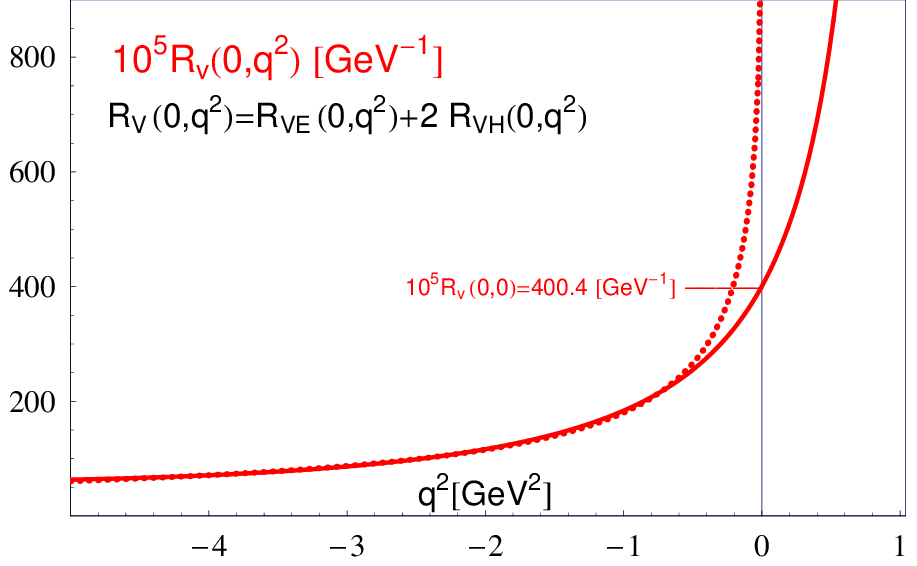}    &  \includegraphics[height=5cm]{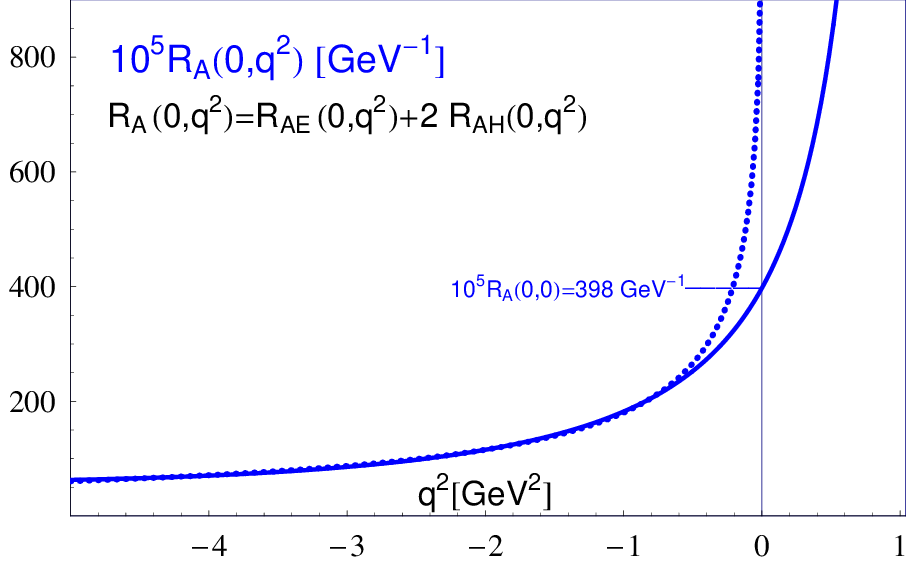}\\
  \end{tabular}
\caption{\label{Plots:H}
  Solid lines are the fits. For $R(0,q^2)$, dashed lines present the results of direct calculations. 
\label{Plot:H}}
\end{center}
\end{figure}

Fig.~\ref{Plots:H} shows our numerical predictions for the NF form factors corresponding to the central
values of all parameters, for the discussion we refer to \cite{bbm2023}. As reported in \cite{bbm2023},
the accuracy of the predictions for the form factors
depend sizebly on $\lambda_{B_s}$. However, for a given value of $\lambda_{B_s}$, the form factors $H_i(k'^2,k^2)$
may be calculated with an accuracy around 10\%. 

\subsection{NF charm vs top}
The effect of factorizable charming loops may be conveniently described as a process-independent
but $q^2$-dependent correction to the Wilson coefficient $C_9$, Eq.~(\ref{C9eff}), with $\delta^{\rm F}C_9(q^2)>0$.  

One may in principle describe also NF charming-loop contribution as a correction to $C_9$;
in this case, however, the correction explodes at small $q^2$. So it is more natural 
to describe the effect of NF charm in $B_s\to\gamma ll$ as additions to the Wilson coefficient $C_{7}$
related to different $A_i(q^2)$ ($i=A,V$) in Eq.~(\ref{fullamplitude}):
\begin{eqnarray}
A_V(q^2): && C_{7}\to C_{7}+\Delta^{\rm NF}_V C_{7}(q^2),\nonumber\\
A_A(q^2): && C_{7}\to C_{7}+\Delta^{\rm NF}_A C_{7}(q^2),
\end{eqnarray}
with the relative correction 
\begin{eqnarray}
\label{deltaC7}
\delta^{\rm NF}_{i}C_{7}(q^2)=\frac{\Delta^{\rm NF}_{i}C_{7}(q^2)}{C_{7}}=8\pi^2
\,Q_s Q_c \frac{C_2}{C_{7}}\frac{1}{m_b}
\frac{\tilde H^{\rm NF}_{i}(q^2,0)+\tilde H^{\rm NF}_{i}(0,q^2)}{F_{Ti}(q^2,0)+F_{Ti}(0,q^2)}, \qquad i=A,V.
\end{eqnarray}
The form factors $F_{Ti}(k'^2,k^2)$ have been evaluated using the 2DAs $\phi_{\pm}$ belonging to the same set of the 
distribution ampltides as the 3DAs (\ref{phi3}) and (\ref{phi4}), see details in \cite{bbm2023}. 
Convenient parametrizations for the form factors $F_{Ti}(k'^2,k^2)$ in a broad range of their momenta are given in Appendix D. 

\begin{figure}[t!]
\begin{center}
\begin{tabular}{cc}   
  \includegraphics[height=4.8cm]{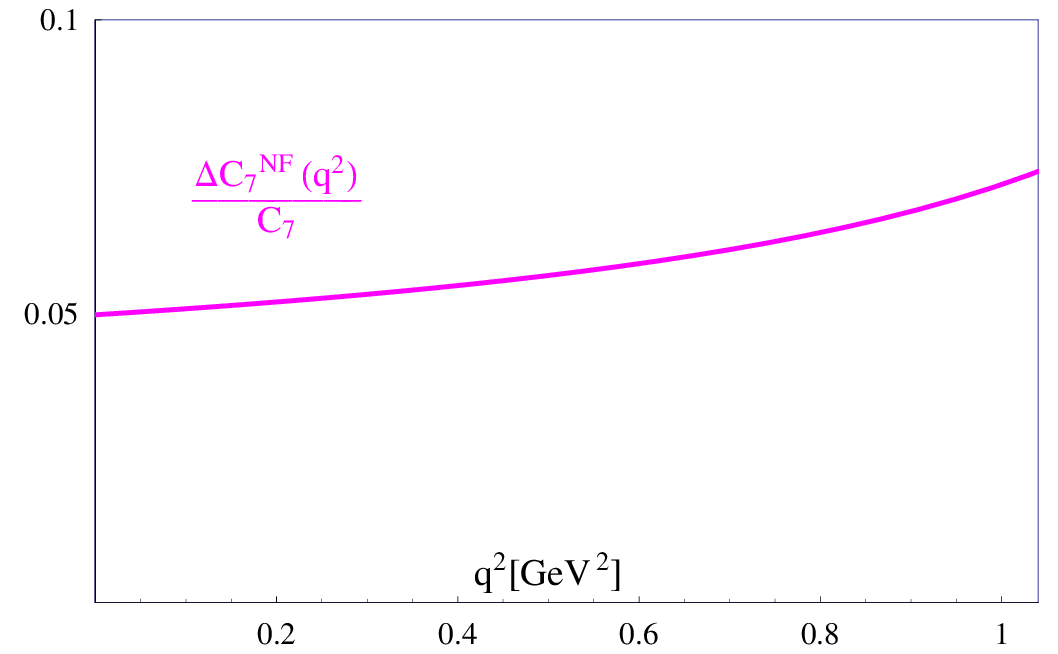}     &  \includegraphics[height=4.8cm]{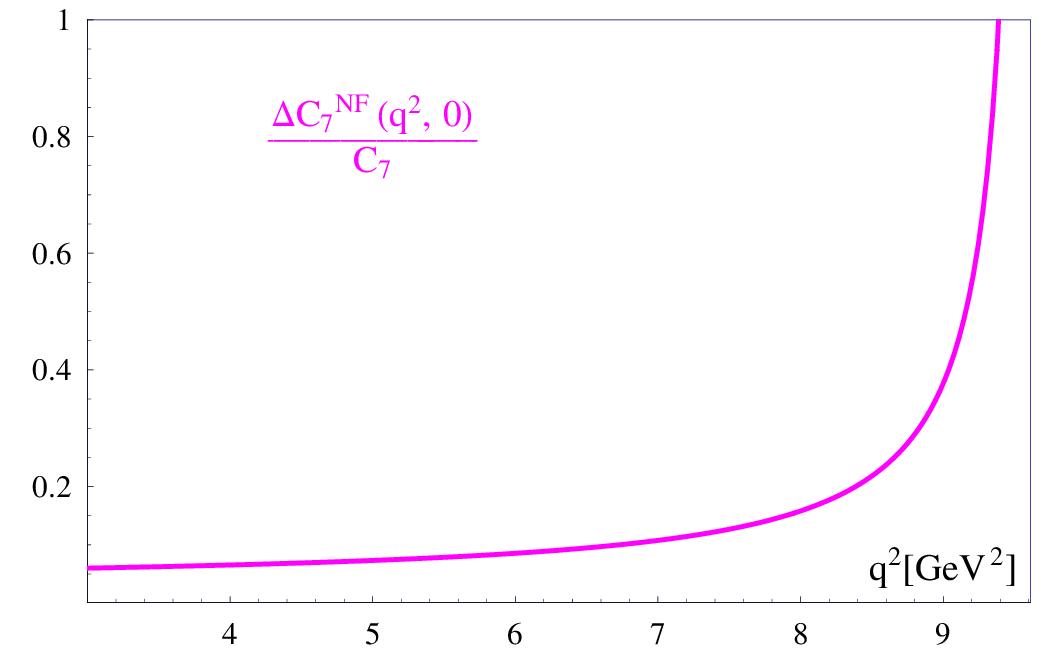}\\
  (a) & (b)          \\
  \end{tabular}   
\caption{ (a) The relative NF correction $\delta^{\rm NF}_{V}C_{7}(q^2)$ given by Eq.~(\ref{deltaC7}).
  (b) $\delta^{\rm NF} C_7(q^2,0)$ given by Eq.~(\ref{deltaC7}) which dominates in 
  $\delta^{\rm NF}_{V}C_{7}(q^2)$ for $q^2> 3 {\rm GeV}^2$
\label{Plot:deltaC7}}
\end{center}
\end{figure}

The Wilson coefficients $C_2$ and $C_7$ have opposite signs,
$\tilde H^{\rm NF}_{i}(q^2,0)$ and $\tilde H^{\rm NF}_{i}(0,q^2)$ as well as  
$F_{Ti}(q^2,0)$ and $F_{Ti}(0,q^2)$ are positive \cite{mnk2018}.
So, the relative correction $\delta^{\rm NF}_{i}C_{7}$ is found to be positive
\begin{eqnarray}
\delta^{\rm NF}_{i}C_{7}(q^2)>0. 
\end{eqnarray}
Numerically, $\delta^{\rm NF}_{A}C_{7}(q^2)\simeq \delta^{\rm NF}_{V}C_{7}(q^2)$. 
The form factors $H_{i}(q^2,0)$ are predicted in the region $q^2<M^2_{J/\psi}$, whereas
$H_{i}(0,q^2)$ is predicted in the region $q^2<4M^2_K$ [Recall that at $q^2>4M^2_{K}$,
$H_{i}(0,q^2)$ have imaginary part]. In principle, one can model $H_{i}(0,q^2)$ for $q^2>4M^2_{K}$,
but this interesting problem is beyond the scope of this paper. So, Fig.~\ref{Plot:deltaC7}(a) 
presents $\delta^{\rm NF}_{i}C_{7}(q^2)$ in the range $0<q^2<4M^2_{K}$, where our predictions
are less model dependent. On the other hand, as the analysis of \cite{mnk2018} has shown, for $q^2\ge 3$ GeV$^2$
(i.e., far above $M_\phi^2$), 
the contribution of the amplitude $H^{\rm NF}_i(0,q^2)$ turns out to be much suppressed
compared to $H^{\rm NF}_i(q^2,0)$. The same occurs for the form factor $F_{Ti}$: in this range of $q^2$,
$F_{Ti}(q^2,0)\gg F_{Ti}(0,q^2)$. Therefore the contribution of $H^{\rm NF}_i(0,q^2)$ in the numerator 
and the contribution of $F_{Ti}(0,q^2)$ in the denominator of (\ref{deltaC7}) may be neglected
(one however might need to be careful as both $H^{\rm NF}_i(0,q^2)$ and $H^{\rm NF}_i(0,q^2)$ have imaginary parts at $q^2>4M_K^2$). 
Then the main contribution to $\delta^{\rm NF} C_7(q^2)/C_7$ in this range of $q^2$ is expected to come from the ratio
$\frac{H^{\rm NF}_{i}(q^2,0)}{F_{Ti}(q^2,0)}$. This contribution is denoted as
\begin{eqnarray}
  \label{deltaC70}
  \delta^{\rm NF}_i C_7(q^2,0)=\Delta^{\rm NF}_i C_7(q^2,0)/C_7, 
\end{eqnarray}
and is shown in Fig.~\ref{Plot:deltaC7}(b) for $i=V$.
Closing this Section, we would like to emphasize that factorizable and nonfactorizable contributions
of the charming loops, $H_i^{\rm F}$ and $H_i^{\rm NF}$, have {\it opposite} signs.

\section{\label{sec:conclusions} Discussion and conclusions}
This paper extended the theoretical approach to NF charming loops in FCNC $B_s$ decays recently
formulated in \cite{bbm2023} and for the first time reports the results for NF charm in $B_s\to\gamma ll$ decays: 

\vspace{0.2cm}
\noindent
(i) We derived analytical expressions for the form factors $H_i^{\rm NF}(k'^2,k^2)$, $i=A,V$,
describing NF contribution of charming loops to the amplitude of the $B_s$ meson transition
into two virtual photons (the first argument, $k'^2$
corresponds to the momentum emitted from the charming loop, whereas the second argument, $k^2$,
corresponds to the momentum emitted by the valence $s$-quark of the $B_s$-meson).
These expressions may be written in the form 
\begin{eqnarray}
H_i^{\rm NF}(k'^2,k^2)=2C_2 Q_s Q_c \left[\lambda_E^2 R_{iE}(k'^2,k^2)+\lambda_H^2 R_{iH}(k'^2,k^2)\right]\qquad i=A,V,3.
\end{eqnarray}
Since an approximate relation $\lambda_H^2\simeq 2\lambda_E^2$ is expected, the linear combination 
\begin{eqnarray}
R_{i}(k'^2,k^2)= 2 R_{iH}(k'^2,k^2) +R_{iE}(k'^2,k^2) 
\end{eqnarray}
is approprate for the description of NF charming loops in $B_s$ decays such that 
\begin{eqnarray}
H_i^{\rm NF}(k'^2,k^2)=2C_2 Q_s Q_c\lambda_E^2 R_{i}(k'^2,k^2).
\end{eqnarray}
We emphasize that according to our analysis, $R_{i}(k'^2,k^2) > 0$ in the region $k'^2<M_{J/\psi}^2$ and $k^2<M_{\phi}^2$
and thus $H_i^{\rm NF}(k'^2,k^2)$ is {\it positive} in this region. Recall that
the factorizable contribution $H_i^{\rm F}(k'^2,k^2)$ is {\it negative}.

\vspace{.2cm}\noindent
(ii) The analytic expressions allow one to calculate the form factors $R_i(k'^2,k^2)$ in a broad range
$k'^2 < 4m_c^2$ and $k^2 < 0$. 
However, calculations based on finite-order QCD diagrams are not expected to provide good description
of the physical hadron amplitudes near quark thresholds 
(for instance, the calculated form factors exhibit steep rise at $k^2\to 0$ which is unphysical,
as the nearest meson pole lies at $k^2=M_{\phi}^2$ and the two-meson threshold lies at $k^2=4M_K^2$). 
So we pursue the following strategy: 
We make use of the results of our calculation in the rectangular region
$0 <k'^2({\rm GeV}^2) < 4$ and $-5 <k^2({\rm GeV}^2) < -0.6$ (i.e., sufficiently far from quark thresholds)
and {\it interpolate} them by a simple analytic formula depending on $k'^2$ and $k^2$, which
takes into account the presence of the poles at $k'^2=M_{J/\psi}^2$ and $k^2=M_{\phi}^2$.
Numerical parameters in this formula are obtained by the fit to the results of our calculations
and interpolate them with a 2\% accuracy in the rectangular region mentioned above. 
The corresponding easy-to-use fit formulas for $R_i(k'^2,k^2)$ are presented in Appendix \ref{Appendix:C}. 

\vspace{.2cm}
\noindent
(iii) Since the interpolating formulas exhibit the correct location of the lowest hadron singularities,
i.e., poles at $k'^2=M_{J/\psi}^2$ and at $k^2=M_{\phi}^2$, our fit formulas are expected to provide reliable theoretical
predictions for the form factors in a broader range $0 <k'^2 < M_{J/\psi}^2$ and $-5\,{\rm GeV}^2 <k'^2 < M_{\phi}^2$.
Fig.~\ref{Plot:H} shows $R_i(0,q^2)$ and $R_i(q^2,0)$ related to $B_s\to\gamma ll$ decays.

\vspace{.2cm}
\noindent
(iv) The contribution of factorizable charm in $B_s\to\gamma ll$ decay may be treated as the $q^2$-dependent
correction to the Wilson coefficient $C_9$, such that $\Delta^{\rm F} C_9(q^2)/C_9 > 0$ at $q^2<M_{J/\psi}^2$.
At the same time, the contribution of nonfactorizable charm in $B_s\to\gamma ll$ decay
may be conveniently treated as the $q^2$-correction to the Wilson coefficient $C_7$, such that
$\Delta^{\rm NF}C_7(q^2)/C_7 > 0$ at $q^2<4M_K^2$ (at higher values of $q^2$ the physical NF charming loop has 
imaginary part). Fig.~\ref{Plot:deltaC7} presents our prediction for these quantities.\footnote{
One may in principle interpret NF charm correction as $\Delta^{\rm NF}C_9$ ---
instead of interpreting it as $\Delta^{\rm NF}C_7$ correction as we do here. 
In this case, $\Delta^{\rm NF}C_9(q^2)/C_9$ comes out {\it negative} and explodes at small $q^2$.
So we do not find this possibility to be attractive.}

\vspace{.2cm}\noindent
(v) Our numerical results for the form factors $H^{\rm NF}_i(k'^2,k^2)$ depend sizeably on the precise value of the
parameter $\lambda_{B_s}$. In this respect we see the same picture as for the $B_s\to\gamma\gamma$ decay,
Fig.~7 in \cite{bbm2023}. And, similar to the form factors for $B_s\to\gamma\gamma$ decay, for a fixed value
of $\lambda_{B_s}$, $H^{\rm NF}_i(k'^2,k^2)$ may be calculated with about 10\% accuracy. 

\vspace{.2cm}
It might be useful to recall that the $B_s\to\gamma ll$ decay amplitude receives contributions from the
weak-annihilation type diagrams \cite{wa1,wa2,wa3,wang2024}. The weak-annihilation mechanism differs very much from
the mechanism discussed in this paper and is therefore beyond the scope of our interest here. However,
weak-annihilation diagrams should be taken into account in a complete analysis of $B_s\to\gamma ll$ decays. 

\vspace{.5cm}\noindent 
{\it Acknowledgments.} 
We are pleased to express our gratitude to Yu-Ming Wang for his illuminating remarks
and comments and to Otto Nachtmann, Hagop Sazdjian, and Silvano Simula for valuable discussions.
D.~M. gratefully acknowledges participation at the Erwin Schr\"odinger Institute (ESI) thematic program
``Quantum Field Theory at the Frontiers of the Strong Interaction'' which promoted a deeper understanding
of the problems discussed in this paper.
The research was carried out within the framework of the program ``Particle Physics and Cosmology'' 
of the National Center for Physics and Mathematics.


\appendix
\section{Constraints on the transition form factors  \label{Appendix_constraints}}
\renewcommand\theequation{A.\arabic{equation}}
We present here a discussion of the constraints imposed by the electromagnetic gauge invariance on the 
$\langle  \gamma^*|\bar q O_i b|\bar B_q(p)\rangle$ transition amplitudes induced by the vector, axial-vector, tensor, and pseudotensor weak currents. 
This discussion extends the discussion of \cite{mk2003} and includes also the case when the real photon is emitted from the 
FCNC $b\to q$ vertex. The corresponding form factors are functions of two variables, $k'^2$ and $k^2$, 
where $k'$ is the momentum of the weak $b\to q$ current, 
and $k$ is the momentum of the electromagnetic current, $p=k+k'$. 
Gauge invariance provides constraints on some of the form factors describing the transition 
of $B_q$ to the real photon emitted directly from the quark line, i.e. for the form factors at $k^2=0$. 

These form factors fully determine the amplitudes of the FCNC $B$-decays into leptons in the final state. 
For instance, the four-lepton decay of the $B$ meson requires the form factors $f_i(k'^2,k'^2)$ for $0<k^2,k'^2<M_B^2$. 
For the case of the $B\to \gamma l^+l^-$ transition one needs the form factors $f_i(k'^2=q^2,k'^2=0)$ and $f_i(k'^2=0,k'^2=q^2)$, where
$q$ is the momentum of the $l^+l^-$ pair. 

\subsection{Form factors of the vector weak current}
In case of the vector FCNC current, the gauge-invariant amplitude contains one form factor $g(k'^2,k^2)$:
\begin{eqnarray}
\label{vector}
T_{\alpha,\mu}=i\int dx e^{i k x}
\langle 0| T\left\{ j^{\rm e.m.}_\alpha(x), \bar q\gamma_\mu b(0)\right\}|\bar B_q(p)\rangle=
e\,\epsilon_{\mu\alpha k'k} 2g(k'^2,k^2). 
\end{eqnarray}
The amplitude is automatically transverse and is free of the kinematic singularities so no constraints on $g(k'^2,k^2)$ emerge. 

\subsection{Form factors of the axial-vector weak current}
For the axial-vector current, the corresponding amplitude has three independent gauge-invariant structures and three form factors, and 
in addition has the contact term which is fully determined by the conservation of the electromagnetic current, $\partial^\mu j_\mu^{e.m.}=0$: 
\begin{eqnarray}
\label{axial-vector}
T^5_{\alpha,\mu}&=&i\int dx e^{i k x}
\langle 0| T\left\{ j^{\rm e.m.}_\alpha(x), \bar q\gamma_\mu\gamma_5 b(0)\right\}|\bar B_q(p)\rangle\nonumber
\\
&=&
i e\,\left(g_{\mu\alpha}-\frac{k_\alpha k_{\mu}}{k^2}\right)f(k'^2,k^2)
+
ie\,\left(k'_{\alpha}-\frac{kk'}{k^2}k_\alpha\right)\bigg[
p_{\mu} a_1(k'^2,k^2) + k_{\mu} a_2(k'^2,k^2)\bigg]
+iQ_{B_q}e\,f_{B_q}\frac{k_\alpha p_\mu}{k^2}.  
\end{eqnarray} 
Here $Q_{\bar B_q}=Q_b-Q_q$ is the electric charge of the $\bar B_q$ meson and $f_{\bar B_q}>0$ is defined according to  
\begin{eqnarray}
\langle 0|\bar q\gamma_\mu\gamma_5 b|\bar B_q(p)\rangle =if_{\bar B_q}p_\mu. 
\end{eqnarray}
The kinematical singularity in the projectors at $k^2=0$ should not be the singularity of the amplitude, and therefore gauge invariance 
yields the following relation between the form factors at $k^2=0$:  
\begin{eqnarray}
\label{constraint1}
\left[f+(k'k)a_2\right]_{k^2=0}=0,\qquad a_1(k'^2,k^2=0)=Q_{\bar B_q} f_{\bar B_q}. 
\end{eqnarray} 
For the neutral $\bar B_{d,s}$ mesons, the contact term is absent and therefore the form factor $a_1$ should vanish at 
$k^2=0$, $a_1(k'^2,k^2=0)=0$. This relation is fulfilled automatically, as the two contributions, corresponding to the  
the photon emission from the valence $b$-quark and from the valence $s,d$-quark cancel each other at $k^2=0$. 

The amplitude of the transition to the real photon is described by a single form factor 
\begin{eqnarray}
\label{axial-vector1}
\langle \gamma(k)|\bar q \gamma_\mu\gamma_5 b|\bar B_q(p)\rangle=
-ie\varepsilon^{\alpha}(k)\left( g_{\mu\alpha} k'k - k'_\alpha k_\mu\right)a_2(k'^2,k^2=0). 
\end{eqnarray}

\subsection{Form factors of the tensor weak current}
The transition amplitudes induced by the tensor weak current can be decomposed in the Lorentz structures transverse 
with respect to $k_\alpha$:  
\begin{eqnarray}
\label{tensor}
T_{\alpha,\mu\nu}&=&i\int dx e^{i k x} \langle 0|\left\{T j_\alpha^{e.m.}(x), \bar q \sigma_{\mu\nu} b(0)\right\}|\bar B_q(p)\rangle \nonumber\\
&=&
ie\,\left(\epsilon_{\mu\nu\alpha p}-\frac{k_\alpha}{k^2}\epsilon_{\mu\nu k p}\right)g_1(k'^2,k^2)
+ie\,\epsilon_{\mu\nu\alpha k}g_2(k'^2,k^2)
+ie\,\left(p_\alpha-\frac{pk}{k^2}k_\alpha\right)
\epsilon_{\mu\nu k' k}g_0(k'^2,k^2),
\end{eqnarray}
The contact terms are absent in this amplitude as well as in the amplitude of the pseudotensor current. 
The kinematic singularity of the projectors at $k^2=0$ should not be the singularity of the amplitude, therefore 
\begin{eqnarray}
\label{constraint_tensor}
\left[g_1-(kp)g_0\right]_{k^2=0}=0. 
\end{eqnarray}
Multiplying (\ref{tensor}) by $k'_\nu$, we obtain the penguin transition amplitude 
\begin{eqnarray}
\label{tensor_penguin}
i\int dx e^{i k x} 
\langle 0|\left\{T j_\alpha^{e.m.}(x), \bar q \sigma_{\mu\nu} k'^\nu b(0)\right\}|\bar B_q(p)\rangle =ie\,\epsilon_{\mu \alpha k p}(g_1+g_2). 
\end{eqnarray}
Notice that the penguin amplitude contains only one combination of the form factors. Nevertheless, 
the requirement of the regularity of the amplitude (\ref{tensor}) yields the constraint (\ref{constraint_tensor}). 

\subsection{Form factors of the pseudotensor weak current}
The transition amplitude of the pseudotensor weak current is given in terms of the same form factors as the amplitude (\ref{tensor}), 
and, similar to (\ref{tensor}), contains no contact terms: 
\begin{eqnarray}
\label{pseudotensor}
T^5_{\alpha,\mu\nu}&=&i\int dx e^{i k x} 
\langle 0|\left\{T j_\alpha^{e.m.}(x), \bar q \sigma_{\mu\nu} \gamma_5 b(0)\right\}|\bar B_q(p)\rangle \\
&=& \nonumber
\bigg[
\bigg(g_{\alpha\nu}-\frac{k_\alpha k_{\nu}}{k^2}\bigg)p_\mu-
\bigg(g_{\alpha\mu}-\frac{k_\alpha k_{\mu}}{k^2}\bigg)p_\nu\bigg]e\,g_1
+
(g_{\alpha\nu}k_\mu-g_{\alpha\mu}k_\nu)e\,g_2
+
\bigg(p_{\alpha}-\frac{k\cdot p}{k^2}k_\alpha\bigg)(k_\mu p_\nu-p_\nu k_\mu)e\,g_0. 
\end{eqnarray}
The kinematical singularity in the projectors at $k^2=0$ should cancel in the amplitude, again leading to the constraint 
Eq.~(\ref{constraint_tensor}). 

For the penguin pseudotensor amplitude we then obtain
\begin{eqnarray}
\label{pseudotensor_penguin} 
&&
i\int dx e^{i k x} 
\langle 0|\left\{T j_\alpha^{e.m.}(x), \bar q \sigma_{\mu\nu} \gamma_5 k'^\nu b(0)\right\}|\bar B_q(p)\rangle \nonumber\\
&&=
e\,\left(k'_\alpha k_\mu-g_{\alpha\mu} kk'\right)\left\{g_1+g_2+\frac{k'^2}{kk'}g_1\right\}
+
e\,\left(k'_\alpha-\frac{kk'}{k^2}k_\alpha\right)\left(k_\mu-\frac{kk'}{k'^2}k'_\mu\right)\frac{k'^2}{kk'}\left\{kk' g_0-g_1\right\}. 
\end{eqnarray}
Notice that the contribution of the second Lorentz structure in (\ref{pseudotensor_penguin}) vanishes both for 
$k^2=0$ (because of the constraint Eq.~(\ref{constraint_tensor}): at $k'^2=0$, $kp=kk'$) and for $k'^2=0$. 
However, it does not vanish for both $k^2,k'^2\ne 0$; therefore, the second Lorentz structure contributes to the 
amplitude of the four-lepton decays. 

We can now build the bridge to the form factors which describe the real photon emission by the valence quarks 
defined in Eq.~(\ref{ffs}): denoting the momentum of the $l^+l^-$ pair as $q$, i.e., setting $k^2=0$ 
and replacing $k'^2\to q^2$, we obtain the form factors in Eq.~(\ref{ffs}) through the form factors 
$g,a_2,g_2,g_1(k'^2=q^2,k^2=0)$: 
\begin{eqnarray}
\label{rel1}
&&
F_V(q^2,0)=2M_{B} g(q^2,0),\qquad  
F_A(q^2,0)=-M_{B} a_2(q^2,0), \\ 
\label{rel2}
&&
F_{TV}(q^2,0)=-\left[g_2(q^2,0)+g_1(q^2,0)\right],\qquad  
F_{TA}(q^2,0)=-\left[g_2(q^2,0)+\frac{M_B^2+q^2}{M_B^2-q^2}g_1(q^2,0)\right].
\end{eqnarray}
The form factors describing the real photon emission from the penguin, are obtained by setting $k'^2=0$ and replacing 
$k^2\to q^2$ in the form factors $g_{1,2}(k'^2,k^2)$:
\begin{eqnarray}
\label{rel3}
F_{TV}(0,q^2)=F_{TA}(0,q^2)=-[g_2(0,q^2)+g_1(0,q^2)].
\end{eqnarray}


\renewcommand\theequation{B.\arabic{equation}}
\section{Derivation of $H^{\rm NF}_{\mu\alpha}$}
\label{sec:appB}
\renewcommand\theequation{B.\arabic{equation}}
Here we provide the derivation of Eq.~(\ref{HNF1}).
Our starting point is the matrix element 
\begin{eqnarray}
  H_{\mu\alpha}(k',k)=i\int dz e^{i k' z}\langle0|T\{e Q_c\bar c\gamma_\mu c(z),
  e Q_s \bar s \gamma_\alpha s(0)\}|B_s(p)\rangle, \qquad p=k'+k, 
\end{eqnarray}
where quark operators are Heisenberg operators in the SM, i.e. the corresponding $S$-matrix
includes weak and strong interactions. The nonfactorizable contribution is related to the
octet-octet part of the weak Hamiltonian and requires the emission of at least one soft gluon 
from the charm-quark loop: 
\begin{eqnarray}
\label{A3}
H^{\rm NF}_{\mu\alpha}(k',k) = i \int d z e^{i k' z}
\langle 0|T\lbrace\bar e Q_c c(z)\gamma_{\mu}c(z), i\int dy\,
L^{b\to s\bar cc[8\times8]}_{\text{weak}}(y), i\int dx\, L_{Gcc}(x),
eQ_s\,\bar s(0)\gamma_{\alpha}s(0)\rbrace| \bar B_s(p)\rangle.
\end{eqnarray}
We place $L_{\rm weak}$ at $y=0$ by shifting coordinates
of all operators through the translation
${\cal O} (x)= e^{i\hat P y} {\cal O}(x-y) e^{-iy\hat P}$.  
Using the relations 
$\langle 0|e^{i(\hat P y)} = \langle 0|$ and $e^{-i(y\hat P)} | B_s(p)\rangle=e^{-i(py)} | \bar B_s(p)\rangle$,
and changing the variables $x-y\to x$, $z-y\to z$, $y\to -y$, we find 
\begin{eqnarray}
  \label{A4}
H^{\rm NF}_{\mu\alpha}(k',k) = i^3\, e^2\,Q_c Q_s\int dx dy dz \: e^{i k' z+i k y}
\langle 0|T\lbrace\bar c(z)\gamma_{\mu}c(z),  L^{b\to s\bar cc[8\times8]}_{\text{weak}}(0),
L_{Gcc}(x), \bar s(y)\gamma_{\eta}s(y)\rbrace| \bar B_s(p)\rangle.  
\end{eqnarray}
Taking into account that $L^{b\to s\bar cc[8\times8]}_{\rm weak}=-H^{b\to s\bar cc[8\times8]}_{\rm weak}$
[the latter given by Eq.~(\ref{Hb2scc})],
we obtain
\begin{eqnarray}
\label{A5}
H^{\rm NF}_{\mu\alpha}(k',k)& =& 
-2C_2\,\frac{G_F}{\sqrt{2}} V_{cb}V^*_{cs}e^2\,Q_c Q_s\, i\int dx dy d z\: e^{i k' z+i k y}
\langle 0 |T\lbrace\bar c(z)\gamma_{\mu}c(z),
\bar c(0) \gamma_{\beta}(1-\gamma_5)t^a c(0), \bar c(x) \gamma_\nu t^b c(x)|0 \rangle
\nonumber\\
&&\hspace{4cm}\times
\langle 0|T\left\{\bar s(y)\gamma_{\alpha}s(y),
\bar s(0)\gamma_{\beta}(1-\gamma_5)t^a b(0),B^b_\nu(x)\right\}| \bar B_s(p)\rangle
\end{eqnarray}
It is convenient to insert under the integral (\ref{A5}) the identity
\begin{eqnarray}
  \label{A6}
B^b_\nu(x)=  \frac{1}{(2\pi)^4} \int d\kappa dx' B^b_\nu(x' ) e^{i\kappa (x-x')}.
\end{eqnarray}
This allows us to isolate the contribution of the charm-quark loop $\Gamma^{\beta\nu\mu(ab)}_{cc}(\kappa ,q)$: 
\begin{eqnarray}
\label{A7}
H^{\rm NF}_{\mu\alpha}(k',k)& =& -2C_2\,\frac{G_F}{\sqrt{2}} V_{cb}V^*_{cs} e^2 Q_c Q_s\frac{i}{(2\pi)^4}\int  dy \, e^{i k y} \;
d \kappa e^{-i\kappa x'}
\langle 0|T\lbrace\bar s(0)\gamma_{\beta}(1-\gamma_5) b(0), B^b_\nu(x'), \bar s(y)\gamma_{\alpha}s(y)\rbrace| B_s(p)\rangle
\nonumber
\\
&&\times\underbrace{
  \int dx dz \, e^{i q z+i \kappa x}
  \langle 0 |T\lbrace\bar c(z)\gamma_{\mu}c(z),\bar c(0) \gamma_{\beta} (1-\gamma_5)t^a c(0),
  \bar c(x) \gamma_\nu t^b  c(x)\rbrace|0 \rangle
}_{\Gamma^{\beta\nu\mu(ab)}_{cc}(\kappa ,q)}. 
\end{eqnarray} 
Using momentum representation for the $s$-quark propagator
\begin{eqnarray}
  \langle 0|T\{s(y)\bar s(0) \}|0\rangle =\frac{1}{(2\pi)^4i}\int d\tilde k\,
  e^{-iky}\frac{\tilde{\slashed{k}}+m_s}{m_s^2-\tilde k^2-i0},
\end{eqnarray}
we obtain 
\begin{eqnarray}
  H^{\rm NF}_{\mu\alpha}(k',k)&=&-2C_2\,\frac{G_F}{\sqrt{2}} V_{cb}V^*_{cs}e^2 Q_c Q_s\\
&&\times\underbrace{\frac{1}{(2\pi)^8}\int{d\tilde k}dy e^{-i(k-q')y} dx d\kappa  e^{-i\kappa x}
\Gamma_{cc}^{\beta\nu\mu(ab)}(\kappa ,q)
\langle 0|\bar s(y)\gamma^\alpha\frac{\tilde{\slashed  k}+m_s}{m_s^2-\tilde k^2}
\gamma^\beta(1-\gamma^5)t^a B^b_{\nu}(x)b(0)|\bar B_s(p)\rangle}_{\bar H^{\rm NF}_{\mu\alpha}}.\nonumber
\end{eqnarray}

\section{Numerical results for the form factors $R_{A,V}(k'^2,k^2)$\label{Appendix:C}}
\renewcommand\theequation{C.\arabic{equation}}
We have calculated the form factors $R_i(k'^2,k^2)$ in the region $-5 <k^2({\rm GeV}^2) < 0$ and $0 <k'^2({\rm GeV}^2) < 4m_c^2$.
However, calculations based on finite-order QCD diagrams cannot be trusted near quark thresholds
(for instance, the calculated form factors exhibit steep rise at $k^2\to 0$ which is unphysical,
as the nearest meson pole lies at $k^2=M_{\phi}^2$ and the two-meson threshold lies at $k^2=4M_K^2$). 
So we pursue the following strategy: 
We make use of the results of our calculation in the restricted rectangular region $-5 <k^2(GeV^2) < -0.6$ and $0 <k'^2(GeV^2) < 4$
(i.e., relatively far from quark thresholds) and interpolate them by a simple analytic formula which
takes into account the presence of the poles at $k'^2=M_{J/\psi}^2$ and $k^2=M_{\phi}^2$. Numerical parameters in this formula
are obtained by the fit in the mentioned restricted area. 

For the form factors $H_V$ and $H_A$ we use the following fitting function: 
\begin{eqnarray}
  \label{fitHVNF}
  R_{A,V}(y_1,y_2)&=&\frac{R_{00}}
  {(1-y_1)(1-y_2)\big(1-g_{11} y_1 -g_{12} y_1^2\big)\bigg(1-(a_{20}+a_{21}y_1+a_{22} y_1^2)y_2+(b_{20}+b_{21} y_1+b_{22} y_1^2)y_2^2\bigg)},\nonumber\\
  && 
  y_1\equiv k'^2/M^2_{J/\psi}, \; y_2\equiv k^2/M^2_{\phi}.
\end{eqnarray}
This formula takes into account the correct location of meson poles at $k'^2=M^2_{J/\psi}$ and $k^2=M^2_{\phi}$.
The coefficients in this formula are obtained by {\it interpolation} in the region where our results may be trusted.
The outcome of the fitting procedure is given in Table~\ref{Table:fitHAV}. 

\begin{table}[hb!]
\centering 
\caption{Parameters in (\ref{fitHVNF}) obtained by the interpolation of our numerical results for $R_{V,A}$}
\label{Table:fitHAV}
\begin{tabular}{|l|r|r|r|r|r|r|r|r|r|}
  \hline
              & $R_{00}[{\rm GeV}^{-1}]$ &  $g_{11}$   &  $g_{12}$  & $a_{20}$ & $a_{21}$ & $a_{22 }$       & $b_{20}$ & $b_{21}$ & $b_{22 }$ \\
  \hline
$R_V$  & 400.4   &  $-0.204$    &  0.421    &  0.141   & $-0.041$  & $-0.044$   &  $-0.026$  & 0.006   &  0.005  \\  
$R_A$  & 398.0   &  $-0.154$    &  0.426    &  0.141   & $-0.063$  & $-0.016$   &  $-0.026$  & 0.010   &  0.001  \\
  \hline
\end{tabular}
\end{table}

\section{Numerical results for the form factor $F_{TV}(k'^2,k^2)$ \label{Appendix:D}}
\renewcommand\theequation{D.\arabic{equation}}
The fit formula for the form factor $F_{TV}(k'^2,k^2)$ is obtained by a similar procedure as described in
Appendix \ref{Appendix:C}:

\noindent (i) The numerical results for $F_{TV}(k'^2,k^2)$ are obtained by evaluating the formulas
from Sect.~5~B of \cite{bbm2023} and the 2DAs $\phi_{\pm}$ given in Eqs.~(5.11) and (5.12) of \cite{bbm2023}. 

\noindent (ii) We use the numerical results in the range $0<k'^2({\rm GeV^2})<15$ and $-5<k^2({\rm GeV^2})<-0.6$
(i.e., far below the quark thresholds located at $k'^2=(m_b+m_s)^2$ and $k^2=4m_s^2$)
and interpolate these numerical results using the analytic fit formula 
\begin{eqnarray}
  \label{fitFT}
  F_{TV}(y_1,y_2)&=&\frac{g_{00}}
  {(1-y_1)(1-y_2)\big(1-g_{11} y_1 -g_{12} y_1^2\big)\bigg(1-(a_{20}+a_{21}y_1+a_{22} y_1^2)y_2+(b_{20}+b_{21} y_1+b_{22} y_1^2)y_2^2\bigg)},\nonumber\\
  && 
  y_1\equiv k'^2/M_{B^*_s}^2, \; y_2\equiv k^2/M^2_{\phi}.
\end{eqnarray}
The analytic formula (\ref{fitFT}) reflects the correct location of the physical poles at $k'^2=M_{B^*_s}^2$ and $k^2=M^2_{\phi}$. 
The fit parameters obtained by the interpolation procedure are summarized in Table \ref{Table:fitFT}. The deviation between the
fit formula and the results of the direct calculation are below 2\% in the full range where the interpolation is made. 

\begin{table}[hb!]
\centering 
\caption{Parameters in (\ref{fitFT}) obtained by the interpolation of our numerical results for $F_{TV}$.
}
\label{Table:fitFT}
\begin{tabular}{|l|r|r|r|r|r|r|r|r|r|}
  \hline
      & $g_{00}$ &  $g_{11}$   &  $g_{12}$  & $a_{20}$  & $a_{21}$ & $a_{22 }$  & $b_{20}$ & $b_{21}$   & $b_{22 }$ \\
  \hline
$F_{TV}$ & 0.152   &  $-0.038$  &  $-0.129$  & $-0.197$ & $0.144$ & $0.360$   &  $0.026$  & $-0.021$  & $-0.058$  \\  
  \hline
\end{tabular}
\end{table}


\end{document}